\documentstyle[11pt,epsf]{article}

\newcommand{\newsection}{ \setcounter{equation}{0} \section}

\newcommand{\beq}{\begin{equation}} \newcommand{\eeq}{\end{equation}}
\newcommand{\bea}{\begin{eqnarray}} \newcommand{\eea}{\end{eqnarray}}

  \newcommand
{\Romannumeral}[1]{\uppercase\expandafter{\romannumeral#1}}

\newcommand{\be}{\begin{enumerate}} \newcommand{\ee}{\end{enumerate}}
\newcommand{\bi}{\begin{itemize}} \newcommand{\ei}{\end{itemize}}
\newcommand{\ba}{\begin{array}} \newcommand{\ea}{\end{array}}
\newcommand{\bc}{\begin{center}} \newcommand{\ec}{\end{center}}
\newcommand{\bt}{\begin{tabular}} \newcommand{\et}{\end{tabular}}

\def\lsim{\mathrel{\rlap{\lower4pt\hbox{\hskip1pt$\sim$}}
    \raise1pt\hbox{$<$}}}           
\def\gsim{\mathrel{\rlap{\lower4pt\hbox{\hskip1pt$\sim$}}
    \raise1pt\hbox{$>$}}}           

\newcommand{\half}{\textstyle {1\over2} \displaystyle}    
\newcommand{\third}{\textstyle {1\over3} \displaystyle}   
\newcommand{\Dslash}{{\hbox{D}\kern-0.6em\raise0.15ex\hbox{/}}} 

\begin{document}

\setlength{\oddsidemargin}{0cm} \setlength{\baselineskip}{7mm}

\input epsf

\begin{normalsize}\begin{flushright}
June 2010 \\
\end{flushright}\end{normalsize}

\begin{center}
  
\vspace{20pt}

{\Large \bf Cosmological Density Perturbations }

{\Large \bf with a Scale-Dependent Newton's G }

\vspace{20pt}

{\sl Herbert W. Hamber}
$^{}$\footnote{e-mail address : Herbert.Hamber@aei.mpg.de} 
\\
Max Planck Institute for Gravitational Physics \\
(Albert Einstein Institute) \\
D-14476 Potsdam, Germany\\

and 

{\sl Reiko Toriumi}
$^{}$\footnote{e-mail address : RToriumi@uci.edu}
\\
Department of Physics and Astronomy \\
University of California \\
Irvine, CA 92697-4575, USA \\

\vspace{20pt}

\end{center}

\begin{center} {\bf ABSTRACT } \end{center}

\noindent
We explore possible cosmological consequences of a running Newton's constant $ G ( \Box ) $,
as suggested by the non-trivial ultraviolet fixed point scenario in the quantum field-theoretic
treatment of Einstein gravity with a cosmological constant term.
In particular we focus here on what possible effects the scale-dependent coupling might have
on large scale cosmological density perturbations. 
Starting from a set of manifestly covariant effective field equations
derived earlier, we systematically develop the linear theory of density perturbations for a
non-relativistic, pressure-less fluid.
The result is a modified equation for the matter density contrast, which can be solved and
thus provides an estimate for the growth index parameter $\gamma$ in the presence of a 
running $G$.
We complete our analysis by comparing  the fully relativistic treatment with the corresponding
results for the non-relativistic (Newtonian) case, the latter also with a 
weakly scale dependent $G$.


\vfill


\pagestyle{empty}

\newpage

\pagestyle{plain}

\vskip 10pt
\newsection{Introduction}
\hspace*{\parindent}
\label{sec:intro}

Recent years have seen the development of a bewildering variety of alternative theories
of gravity, in addition to the more traditional alternate theories, which used to include
Brans-Dicke,  tensor-scalar, tensor-vector-scalar, higher derivative, effective quantum
gravity and supergravity theories.
Some of the new additions to the already rather long list include dilaton gravity, 
$f(R)$ and $f(G)$ gravity,  Chern-Simons gravity, conformal gravity, torsion gravity, 
loop quantum gravity, holographic modified gravity, MOG gravity, 
asymmetric brane gravity, massive gravity and minimally modified self-dual gravity,
just to cite a few representative examples. 
All of these theories eventually predict some level of deviation from classical gravity,
which is often parametrized either by a suitable set post-Newtonian parameters, or
more recently by the introduction of a slip function \cite{dam06,dam93}.
The latter has been quite useful in describing deviations from classical GR,
and specifically from the standard $\Lambda CMD$ model,  when analyzing
the latest cosmological CMB, weak lensing, supernovae and galaxy clustering data.

In this paper we will focus on the analysis of departures from GR in the 
growth history of matter perturbations,
within the narrow context of the non-trivial ultraviolet fixed point scenario for Einstein
gravity with a cosmological term.
Thus instead of looking at deviations from GR at very short distances, due
to new interactions such as the ones suggested by string theories \cite{ven02},
we will be considering here infrared effects, which could therefore become
manifest at very large distances.
The classical theory of small density perturbations
is by now well developed in standard textbooks, and the
resulting theoretical predictions for the growth exponents are simple to state,
and well understood.
Except possibly on the very largest scales, where the data so far is still
rather limited, the predictions agree quite well with current astrophysical observations.
Here we will be interested in computing and predicting possible small deviations in the growth
history of matter perturbations, and specifically in the values of the growth exponents,
arising from a very specific scenario, namely a weakly scale-dependent 
gravitational coupling, whose value very gradually increases with distance.

The specific nature of the scenario we will be investigating here is motivated
by the treatment of field-theoretic models of quantum gravity, based on the 
Einstein action with a bare cosmological term.
Its long distance scaling properties are derived from the existence of a non-trivial ultraviolet 
fixed point of the renormalization group in Newton's constant $G$.
The latter is inaccessible by direct perturbation theory in four dimensions, 
and can be shown to radically alter the short- and long-distance behavior
of the theory when compared to more naive expectations.
The renormalization group origin of such fixed points was first discussed in detail by 
Wilson for scalar and self-coupled fermion theories \cite{wil72}.
The general field theoretic methods were later extended and
applied to gravity, where they are now referred to as 
the non-trivial fixed point scenario or asymptotic safety \cite{wei77}.
It is fair to say that so far this is the only field-theoretic approach known 
to work consistently in other not perturbatively renormalizable theories, 
such as the non-linear sigma model.
While perhaps still a bit mundane in the context of gravity, such non-trivial
fixed points are well studied and well understood in statistical field theory,
where they generally describe phase transitions between ordered and
disordered ground states, or between weakly coupled and condensed states.

The paper is organized as follows.
First we recall the effective covariant field equations describing the running
of $G$, and describe the nature of the objects and parameters 
entering the quantum non-local corrections.
We then discuss the zeroth order (in the fluctuations) field equations
and energy-momentum conservation equations
for the standard homogeneous isotropic metric, with a running $G$.
Later we extend the formalism to deal with small metric and matter
perturbations, and derive the relevant field and energy conservation
equations to first order in the perturbations.
After showing the overall consistency of the derived equations, we 
proceed to derive the modified differential equation for the
density contrast $\delta (t)$. 
Later this is re-written, following customary procedures, as a function
of the scale factor as $\delta (a)$.
The resulting differential equation for the density contrast is then solved
and the results for the growth exponents compared to the
standard classical result.
The conclusions provide an interpretation of the theoretical
results and their associated uncertainties vis-\`a-vis
present and future high precision galaxy clustering measurements.

\vskip 40pt
\newsection{Running Newton's Constant ${\bf G(\Box)}$}
\hspace*{\parindent}
\label{sec:gbox}

Originally the running of $G$ was computed either on the lattice directly in four dimensions
\cite{book,hw84,ham00}, or in the continuum within the framework of the background 
field expansion applied to $2+\epsilon$ spacetime dimensions \cite{wei77,eps}
and later using truncation methods applied in 4d \cite{reu98}.
In either case one obtains a momentum dependent $G(k^2)$, which
needs to be eventually re-expressed in a coordinate-independent way, so that it
can be usefully applied to more general problems involving
arbitrary background geometries.

The first step in analyzing the consequences of a running of $G$
is therefore to re-write the expression for $G(k^2)$ in a coordinate-independent
way, either by the use of a non-local Vilkovisky-type effective gravity action \cite{vil84,ven90},
or by the use of a set of consistent effective field equations.
In going from momentum to position space one usually employs 
$k^2 \rightarrow - \Box$, which then gives for
the quantum-mechanical running of the gravitational
coupling the replacement $ G  \;\; \rightarrow \;\; G( \Box ) $.
One then finds that the running of $G$  is given in the vicinity of the UV fixed point by
\beq
G ( \Box ) \, = \, G_0 \left [ \; 1 \, 
+ \, c_0 \left ( { 1 \over \xi^2 \Box  } \right )^{1 / 2 \nu} \, 
+ \, \dots \, \right ] \; ,
\label{eq:grun_box_0}
\eeq
where $\Box \equiv g^{\mu\nu} \nabla_\mu \nabla_\nu$
is the covariant d'Alembertian, and the dots represent higher
order terms in an expansion in $1 / ( \xi^2 \Box ) $.
Current evidence from Euclidean lattice quantum gravity points toward 
$c_0 > 0$ (implying infrared growth) and $\nu \simeq \third $ \cite{ham00}.

Within the quantum-field-theoretic renormalization group treatment, this last quantity arises 
as the integration constant of the Callan-Symanzik renormalization group equations.
One challenging issue therefore, and of great relevance to the physical interpretation of the results, 
is a correct identification of the renormalization group invariant scale $\xi$.
A number of arguments can be given (see below) in support of the suggestion that the
infrared scale $\xi$ (very much analogous to the $\Lambda_{\overline{MS}}$ of QCD) can in 
fact be very large, even cosmological, in the gravity case.
From these arguments one would then infer that the constant $G_0$ can, to a very close 
approximation, be identified with the laboratory value of 
Newton's constant, $ \sqrt{G_0} \sim 1.6 \times 10^{-33} {\rm cm}$.

The appearance of the d'Alembertian $\Box$ in the running of $G$ naturally 
leads to both a non-local effective gravitational action, and a corresponding
set of non-local modified field equations.
Instead of the ordinary Einstein field equations with constant $G$
\beq
R_{\mu\nu} \, - \, \half \, g_{\mu\nu} \, R \, + \, \lambda \, g_{\mu\nu}
\; = \; 8 \pi \, G \, T_{\mu\nu} \; ,
\label{eq:field}
\eeq
one is now lead to consider the modified effective field equations
\beq
R_{\mu\nu} \, - \, \half \, g_{\mu\nu} \, R \, + \, \lambda \, g_{\mu\nu}
\; = \; 8 \pi \, G  ( \Box )  \, T_{\mu\nu} 
\label{eq:field1}
\eeq
with a new non-local term due to the $G(\Box)$.
By being manifestly covariant they still satisfy some
of the basic requirements for a set of consistent field equations
incorporating the running of $G$.
Not unexpectedly though, the new nonlocal equations are much
harder to solve than the original classical field equations for constant $G$.

It is instructive to note, as already pointed out in \cite{hw05},
that the effective non-local field equations of 
Eq.~(\ref{eq:field1}) can be re-cast in a form very similar to the classical
field equations, but with a new source term 
$ {\tilde T_{\mu\nu}} \, = \, \left [ G  ( \Box )  / G_0   \right ] \, T_{\mu\nu}$
defined as the effective, or gravitationally dressed, energy-momentum tensor.
Ultimately the consistency of the effective field equations demands that it
be exactly conserved, in consideration of the contracted Bianchi identity 
satisfied by the Ricci tensor.
In this picture, therefore, the running of $G$ can be viewed 
as contributing to a sort of a vacuum fluid, introduced in order to account for 
the new gravitational vacuum polarization contribution.

More on the technical side,  and mainly due the appearance of a negative
fractional exponent in Eq.~(\ref{eq:grun_box_0}) ,
the covariant operator appearing in the expression for $G(\Box)$
has to be suitably defined by analytic continuation. 
This can be done, for example, by computing $\Box^n$ for positive
integer $n$, and then analytically continuing to $n \rightarrow -1/2\nu$ \cite{hw05}.
Equivalently, $G(\Box)$ can be defined via a suitable regulated parametric integral 
representation \cite{lop07}, such as
\beq
\left ( { 1 \over - \Box (g) + m^2 } \right )^{1/ 2 \nu } 
\, = \, 
{ 1 \over \Gamma ( { 1 \over 2 \nu } ) } \,
\int_0^\infty d \alpha \; \alpha^{  1 / 2 \nu - 1 } \;
e^{  - \alpha \, ( - \Box (g) + m^2 ) } \; .
\label{eq:gbox_exp}
\eeq
As far as the calculations in this paper are concerned, it will not be necessary
to commit oneself to an unduly specific form for the running of $G (\Box )$.
Thus for example, although the lattice gravity results only allow for a non-degenerate phase
for the case $c_0 >0$, it will nevertheless be possible later to have either sign
for the correction in Eq.~(\ref{eq:grun_box_0}), in the sense that the very existence
of a non-trivial ultraviolet fixed point implies in principle the appearance of two
physically distinct phases, each of which might or might not be physically
realized due to issues of non-perturbative stability.
Observation could then be used, in principle, to constrain one or the other choice.
Furthermore, the value of the exponent $\nu$ need not to be specified until the 
very end of the calculation, so that most of the results can be kept general.
\footnote{
A running cosmological constant $\lambda (k) \rightarrow \lambda ( \Box ) $ 
causes a number of mathematical inconsistencies \cite{hw05} within
the manifestly covariant framework, described here
by the effective field equations of Eq.~(\ref{eq:field1}).
Indeed if one assumes that $ \lambda ( \Box ) \sim ( \xi^2 \Box )^{-\sigma} $, where
$\sigma $ is a (positive or negative) power, then for example the infrared regulated expression
in Eq.~(\ref{eq:gbox_exp}) gives no running of $\lambda$, after using
$\nabla_\lambda g_{\mu\nu} =0 $.
This last conclusion is in agreement with the field-theoretic results of the non-trivial renormalization
group fixed point scenario, thereby providing perhaps an independent consistency check.} 

The situation regarding the running of $G$ is perhaps most easily illustrated close 
and above two dimensions, where
the gravitational coupling becomes dimensionless,  $G\sim \Lambda^{2-d}$
with $\Lambda$ the ultraviolet cutoff required to regularize the theory
(a similar and completely parallel line of arguments and results can in fact
be presented for the 4d lattice theory as well, but a discussion of 
renormalization on the lattice ends up being inevitably quite a bit less transparent
\cite{ham00,book}).
There the theory appears perturbatively renormalizable, so
that the full machinery of covariant renormalization and of the renormalization 
group can in principle be applied, following Wilson's dimensional
expansion method, now formulated as a double expansion in
$G$ and $\epsilon=d-2$ \cite{wei77,eps}.
Both here and on the lattice a renormalization of the bare cosmological
constant, besides being gauge-dependent, is also physically meaningless, as 
it can be reabsorbed by a trivial rescaling of the metric;
the latter is needed in order to recover the proper normalization
of the volume term in the path integral, thus avoiding
spurious renormalization effects, as discussed in \cite{book,ham00,eps}.

In momentum space the result corresponding to Eq.~(\ref{eq:grun_box_0}),
and allowing now possibly for either sign in front of the correction, is
\begin{equation}
G(k^2) \; \simeq \; G_0 \, \left [ 
\, 1 \, \pm \, c_0 \, \left ( { 1 \over \xi^2 \, k^2 } \right )^{ 1/ 2 \nu }
\, + \, \dots \right ] \; ,
\label{eq:grun_k} 
\end{equation}
with $c_0$ a positive constant, and $\xi $ the new, genuinely nonperturbative, 
gravity scale.
\footnote{A properly infrared regulated version of the above expression, 
here with the choice of + sign, would read
\beq
G(k^2) \; \simeq \; G_0 \left [ \; 1 \, 
+ \, c_0 \left ( { \xi^{-2} \over k^2 \, + \, \xi^{-2} } \right )^{1 / 2 \nu} \, 
+ \, \dots \; \right ] \; .
\label{eq:grun_k_reg}
\eeq
Then for large distances $r \gg \xi$  the gravitational coupling
no longer exhibits the spurious infrared divergence, but instead approaches
the finite value $G_\infty \simeq ( 1 + c_0 + \dots ) \, G_0 $.}
Consequently the above expression for $G(k^2)$ can be used whenever
the full generality of the manifestly covariant expression in
Eq.~(\ref{eq:grun_box_0}) is not really needed, for example when 
dealing with the Newtonian (non-relativistic) limit.

The choice of $+$ or $-$ sign is ultimately determined from whether one is initially 
to the left (+), or to right (-) of the fixed point $G_0$, in which case
the effective $G(k^2)$ decreases or, respectively, increases as one flows away
from the ultraviolet fixed point towards lower momenta, or larger distances.
Physically the two solutions represent of course gravitational screening ($G<G_0$)
or anti-screening ($G>G_0$).

It is crucial that the quantum correction involves a new physical, renormalization 
group invariant, scale $\xi$, whose value cannot be fixed by a perturbative
calculation, and whose absolute size determines the 
comparison scale for the new non-local quantum effects.
It should therefore be rightfully considered as the gravity analog of the celebrated 
gauge theory scaling violation parameter $\Lambda_{\overline{MS}}$.
In terms of the bare gravitational coupling $G(\Lambda)$ it is given by
\beq
\xi^{-1} = A_{\xi} \cdot \Lambda \, 
\exp \left ( { - \int^{G(\Lambda)} \, {d G' \over \beta (G') } }
\right ) \; ,
\label{eq:m-cont}
\eeq
where $\beta (G)$ is the Callan-Symanzik beta function for $G$
(which can be given explicitly, for example, in the $2+\epsilon$ expansion
to a given loop order, or can be computed on the lattice).
It is then more or less a direct consequence of the renormalization group that
the value of the constant $A_{\xi}$ determines the coefficient 
$c_0$ in Eq.~(\ref{eq:grun_box_0}), $c_0 = 1/ (A_{\xi}^{1/\nu} G_0 )$.
The non-perturbative lattice formulation of quantum gravity 
then allows an explicit and direct computation of 
$A_{\xi}$,  and therefore of the coefficient $c_0$ in $G(\Box)$ \cite{ham00,book}.

Physically it would seem at first that the non-perturbative
scale $\xi$ could take any value, including
a very small one - based on the naive estimate $\xi \sim l_P$ - which would
then of course preclude any observable quantum effects in the foreseeable future.
But a number of recent results for the gravitational Wilson loop on the 
Euclidean lattice at strong coupling, giving an area law, and their subsequent
interpretation in light of the observed large scale semiclassical 
curvature \cite{loo07},
would suggest otherwise, namely that the non-perturbative scale $\xi$
appears in fact to be related to macroscopic curvature.
From astrophysical observation the average curvature on very large scales, 
or, stated in somewhat better terms, the measured
cosmological constant $\lambda$, is very small.
This would then suggest that the new scale $\xi$ can be very large,
even cosmological,
\begin{equation}
{ 1 \over \xi^2 } \; \simeq \; { \lambda \over 3 } 
\label{eq:xi_lambda}
\end{equation} 
which would then give a more concrete quantitative estimate 
for the scale in the $G(\Box)$ of Eq.~(\ref{eq:grun_box_0}), namely
$\xi \sim 1 / \sqrt{\lambda/3} \sim 1.51 \times 10^{28} {\rm cm} $.
Indeed for quantum gravity no other suitable infrared cutoff presents itself, so
that $\lambda$ can almost be considered as the only "natural" candidate to 
take on the role of a (generally covariant) infrared regulator or
graviton mass-like parameter.

Finally let us mention here briefly and for completeness that for a limited number
of metrics it has been possible, after some considerable work,
to find exact solutions, in some regime, to the above effective 
nonlocal field equations.
One such case is the static isotropic metric, where in the limit
$r \gg 2 M G$ one can obtain an explicit solution for the metric
coefficients $A(r) = 1/B(r)$, leading eventually to the rather simple 
result \cite{hw06}
\beq
G \; \rightarrow \; G(r) \; = \; 
G_0 \, \left ( 1 \, + \, 
{ c_0 \over 3 \, \pi } \, m^3 \, r^3 \, \ln \, { 1 \over  m^2 \, r^2 }  
\, + \, \dots \; \right )
\label{eq:g_small_r}
\eeq
with $m \equiv \xi^{-1}$, consistent with a gradual slow increase of $G(r)$ with distance.
\footnote{We have pointed out before that the
result for $G(r)$ is in a number of ways reminiscent of the
analogous QED result (known as the Uehling correction to the Coulomb
potential in atoms)
\begin{equation}
Q \; \rightarrow \; Q(r) \; = \; Q \, \left ( 1 \, + \, 
{\alpha \over 3 \, \pi } \, \ln { 1 \over m^2 \, r^2 } \, + \, \dots \right ) \; .
\label{eq:qed_s}
\end{equation}
In the gravity case the correction is not a log but a power, which is what one
would naively expect from a perturbatively non-renormalizable theory.
In gravity, the infrared cutoff due in QED to the finite physical electron 
mass is naturally replaced by the physical cosmological constant; the magnitude 
of neither one of these two quantities can be predicted by the fundamental theory.}
One amusing aspect of the exact solution in the static isotropic case 
is that no consistent solution can be found 
unless $\nu=1/3$ exactly in four dimensions,
and similarly $\nu =1/(d-1)$ in dimensions $ d \geq 4$ \cite{hw06},
lending further support, and independently of the lattice theory
results, to this particular value for $\nu$ in four dimensions.

\vskip 40pt
\subsection{ (Zeroth Order) Effective Field Equations with ${\bf G(\Box)}$}
\hspace*{\parindent}
\label{sec:field_gbox}

A scale dependent Newton's constant is expected to lead to small modifications
of the standard cosmological solutions to the Einstein field equations.
Here we will summarize what modifications are
expected from the effective field equations on the basis of $G(\Box)$,
as given in Eq.~(\ref{eq:grun_box_0}), which itself originates in
Eq.~(\ref{eq:grun_k}).
The starting point are the quantum effective field equations
of Eq.~(\ref{eq:field1}), 
with $G(\Box)$ defined in Eq.~(\ref{eq:grun_box_0}).
In the Friedmann-Lema\^itre-Robertson-Walker (FLRW) framework these are
applied to the standard homogeneous isotropic metric
\beq
d \tau^2 \; = \;  dt^2 - a^2(t) \left \{ { dr^2 \over 1 - k\,r^2 } 
+ r^2 \, \left ( d\theta^2 + \sin^2 \theta \, d\varphi^2 \right )  \right \} \;\;\;\; k =0, \pm1 \; .
\eeq
In the following we will mainly consider the case $k=0$ (spatially flat universe).
It should be noted that there are in fact {\it two} related quantum contributions to the
effective covariant field equations. 
The first one arises because of the presence of a non-vanishing 
cosmological constant $\lambda \simeq 3 / \xi^2 $, caused by the
non-perturbative vacuum condensate.
As in the case of standard FLRW cosmology, this is expected to be 
the dominant contributions at large times $t$, and gives an exponential
(for $\lambda>0$), or cyclic (for $\lambda < 0$) expansion of the scale factor.
The second contribution arises because of the explicit running of $G (\Box)$ in the 
effective field equations.
The next step therefore is a systematic examination of the nature of
the solutions to the full effective field equations,
with $G ( \Box )$ involving the relevant covariant d'Alembertian operator
\beq
\Box \; = \; g^{\mu\nu} \, \nabla_\mu \nabla_\nu 
\eeq
acting on second rank tensors as in the case of $T_{\mu\nu}$,
\bea
\nabla_{\nu} T_{\alpha\beta} \, = \, \partial_\nu T_{\alpha\beta} 
- \Gamma_{\alpha\nu}^{\lambda} T_{\lambda\beta} 
- \Gamma_{\beta\nu}^{\lambda} T_{\alpha\lambda} \, \equiv \, I_{\nu\alpha\beta}
\nonumber
\eea
\beq 
\nabla_{\mu} \left ( \nabla_{\nu} T_{\alpha\beta} \right )
= \, \partial_\mu I_{\nu\alpha\beta} 
- \Gamma_{\nu\mu}^{\lambda} I_{\lambda\alpha\beta} 
- \Gamma_{\alpha\mu}^{\lambda} I_{\nu\lambda\beta} 
- \Gamma_{\beta\mu}^{\lambda} I_{\nu\alpha\lambda}  \; .
\label{eq:box_on_tensors}
\eeq
and in general requires the calculation of 1920 terms, of which fortunately many vanish 
by symmetry due to specific choice of metric.

To start the process, one assumes for example that $T_{\mu\nu}$ has a perfect fluid form, 
\beq
T_{\mu \nu} = \left [ \, p(t) + \rho(t) \, \right ] u_\mu \, u_\nu + g_{\mu \nu} \, p(t)
\label{eq:tmunu_perf}
\eeq
for which one needs to compute the action of $\Box^n$ on $T_{\mu\nu}$,  and 
then analytically continues the answer to negative fractional values of $n = -1/2 \nu $.
Even in the simplest case, with  $G(\Box)$ acting on a {\it scalar} such as the trace
of the energy-momentum tensor $T_\lambda^{\;\; \lambda}$, one finds for the choice
$ \rho(t) = \rho_0 \, t^{\beta} $ and $ a(t) = a_0 \, t^{\alpha} $
the rather unwieldy expression
\beq
\Box^n \left [ - \rho (t) \right ] \rightarrow 4^n 
\left ( - 1 \right )^{n+1} { \Gamma \left ( {\beta \over 2} + 1 \right ) \, 
\Gamma \left ( {\beta + 3 \, \alpha + 1 \over 2} \right ) \over 
\Gamma \left ( {\beta \over 2} + 1 - n \right ) \, 
\Gamma \left ( {\beta + 3\, \alpha + 1 \over 2} - n \right ) } \, \rho_0 \, t^{\beta - 2 n} \; ,
\eeq
with an integer $n$ later analytically continued to $n \rightarrow - {1 \over 2 \, \nu}$, with
$\nu= \third $.

A more general calculation shows that a non-vanishing pressure contribution is generated in 
the effective field equations, even if one initially assumes a pressureless fluid, $p(t)=0$.
After a somewhat lengthy derivation one obtains for a universe filled with non-relativistic 
matter ($p$=0) the following set of effective Friedmann equations,
\bea
{ k \over a^2 (t) } \, + \,
{ \dot{a}^2 (t) \over a^2 (t) }  
& = & { 8 \pi \, G(t) \over 3 } \, \rho (t) \, + \, { \lambda \over 3 }
\nonumber \\
& = & { 8 \pi \, G_0 \over 3 } \, \left [ \,
1 \, + \, c_t \, ( t / \xi )^{1 / \nu} \, + \, \dots \, \right ]  \, \rho (t)
\, + \, { \lambda \over 3 }
\label{eq:fried_tt}
\eea
for the $tt$ field equation, and
\bea
{ k \over a^2 (t) } \, + \, { \dot{a}^2 (t) \over a^2 (t) }
\, + \, { 2 \, \ddot{a}(t) \over a(t) } 
& = & - \, { 8 \pi \, G_0 \over 3 } \, \left [ \, c_t \, ( t / t_0 )^{1 / \nu} 
\, + \, \dots \, \right ] \, \rho (t) 
\, + \, \lambda
\label{eq:fried_rr}
\eea
for the $rr$ field equation.
In the above expressions the running of $G$ appropriate for 
the RW metric is
\beq
G (t) \, \equiv \, G_0 \left ( 1 + { \delta G(t) \over G_0 } \, \right ) 
\, = \, G_0 \left [ 1 + c_t \, 
\left ( { t \over t_0 } \right )^{1 / \nu} \, + \, \dots \right ] 
\label{eq:grun_t}
\eeq
with $c_t$ of the same order as $c_0$ in Eq.~(\ref{eq:grun_k}), and $t_0 = \xi$ \cite{hw05};
in the quoted reference it was estimated $c_t = 0.450 \; c_0$ for the tensor box operator.
Note that it is the running of $G$ that induces an effective pressure term in the second 
($rr$) equation, corresponding to the presence of a relativistic fluid
due to the vacuum polarization contribution.
One important feature of the new equations is an additional power-law
acceleration contribution, on top of the standard one due to $\lambda$.

\vskip 40pt
\subsection{Introduction of the ${\bf w_{vac}}$ Parameter}
\hspace*{\parindent}
\label{sec:w_vac}

It was noted in \cite{hw05} that the field equations with a running $ G $, Eqs.~(\ref{eq:fried_tt})
and (\ref{eq:fried_rr}), can be recast in an equivalent, but slightly more appealing, 
form by defining a vacuum polarization pressure $p_{vac}$ and density 
$\rho_{vac}$, such that for the FLRW background one has
\beq
\rho_{vac} (t) = {\delta G(t) \over G_0} \, \rho (t)  \;\;\;\;\;\;\;\;\;\;\;\; 
p_{vac} (t) = { 1 \over 3} \, {\delta G(t) \over G_0} \, \rho (t) \; .
\label{eq:rhovac_t}
\eeq
Consequently the source term in the $tt$ field equation can be regarded as
a combination of two density terms
\beq
 \left ( 1 + {\delta G(t) \over G_0} \right ) \, \rho(t)  \equiv  \rho(t) + \rho_{vac} (t) \; ,
\eeq
while the $rr$ equation involves the new vacuum polarization pressure term
\beq
{1 \over 3} \,  {\delta G(t) \over G_0} \, \rho(t)  \equiv  p_{vac} (t) \; .
\eeq
Form this viewpoint, the inclusion of a vacuum polarization contributions in the FLRW 
framework seems to amount to a replacement 
\beq
\rho(t) \rightarrow \rho(t) + \rho_{vac} (t)     
\;\;\;\;\;\;\;\;\;\;\;\;
p(t) \rightarrow p(t) +  p_{vac} (t)
\label{eq:vac}
\eeq
in the original field equations.
Just as one introduces the parameter $w$, describing the matter equation of state, 
\beq
p (t) = w \,  \rho(t)
\label{eq:w_def}
\eeq
with $ w=0 $ for non-relativistic for matter, one can do the same for the remaining  contribution
by setting
\beq
p_{vac} (t) = w_{vac} \; \rho_{vac} (t) \; .
\label{eq:wvac_def}
\eeq
Then in terms of the two $w$ parameters
\beq
\left ( w + w_{vac}  {\delta G(t) \over G_0} \right ) \, \rho (t) = p (t) + p_{vac} (t) 
\eeq
with, according to Eqs.~(\ref{eq:fried_tt}) and (\ref{eq:fried_rr})
and following the results of \cite{hw05},  $w_{vac}= \third $ in a FLRW background.
We should remark here that the calculations of \cite{hw05} also indicate that $w_{vac}= \third $
is obtained {\it generally} for the given class of $G(\Box)$ considered, and is 
not tied therefore to a specific choice of $\nu$, such as $\nu=\third$.

The previous, slightly more compact, notation allows one to re-write the field 
equations for the FLRW background in an equivalent form, which we will describe below.
First we note though that in the following we will restrict our attention mainly to a 
spatially flat geometry, $k=0$.
Furthermore, when dealing with density perturbations we will have to distinguish between
the background, which will involve a background pressure ($\bar p$) and background
density ($\bar \rho$), from the corresponding perturbations which will be denoted here
by $\delta p$ and $\delta \rho$.
Then with this notation and for constant $G_0$,
 the FLRW field equations for the background are written as
\bea
3 \, {{\dot{a}}^2 (t) \over {a}^2 (t)} 
& = & 
8 \pi \, G_0  \, \bar{\rho} (t) + \lambda 
\nonumber \\
{{\dot{a}}^2 (t) \over {a}^2 (t)} + 2\, {\ddot{a} (t) \over a (t)} 
& = &
- 8 \pi \, G_0 \, {\bar p}  (t) + \lambda \; .
\label{eq:fried_0}
\eea
Now in the presence of a running $G(\Box)$,  and in accordance with the results 
of Eqs.~(\ref{eq:fried_tt}) and (\ref{eq:fried_rr}), 
the modified FLRW equations for the background read
\bea
3 \, {{\dot{a}}^2 (t) \over {a}^2 (t)} 
& = & 
8 \pi \, G_0 \left ( 1 + {\delta G(t) \over G_0} \right ) \, \bar{\rho}(t) + \lambda 
\nonumber \\
{{\dot{a}}^2 (t) \over {a}^2 (t)} + 2\, {\ddot{a} (t) \over a (t)} 
& = & 
- 8 \pi \, G_0 \, \left ( w + w_{vac}  {\delta G(t) \over G_0} \right )\, \bar{\rho} (t) + \lambda \; ,
\label{eq:fried_run}
\eea
using the definitions in Eqs.~(\ref{eq:w_def}) and (\ref{eq:wvac_def}),
here with $\bar{p}_{vac} (t) = w_{vac} \,  \bar{\rho}_{vac}(t)$.

We note here that the procedure of defining a $\rho_{vac}$ and a $p_{vac}$ contribution,
arising entirely from quantum vacuum polarization effects, is not necessarily
restricted to the FLRW background metric case \cite{hw05}.
In general one can decompose the full source term in the effective nonlocal
field equations of Eq.~(\ref{eq:field1}), making use of
\beq
G(\Box) = G_0 \, \left ( 1 \, +  {\delta G(\Box) \over G_0} \right ) 
\;\;\;\;\;\;  {\rm with} \;\;\;\;\;
{\delta G(\Box) \over G_0} \equiv c_0 \left ( { 1 \over \xi^2 \Box } \right )^{1 / 2 \nu}  \; ,
\label{eq:grun_box}
\eeq
 as two contributions,
\beq
{ 1 \over G_0 } \, G(\Box) \, T_{\mu\nu}  \, = \, 
\left ( 1 + {\delta G(\Box) \over G_0}  \right ) \, T_{\mu\nu}  \, = \,
T_{\mu\nu}   +  T_{\mu\nu}^{vac} \; .
\label{eq:tmunu_vac}
\eeq
The latter involves the nonlocal part
\footnote{One normally does not include the l.h.s. field equation contribution
$+ \lambda g_{\mu\nu}$ as part of the r.h.s. matter part
$ T_{\mu\nu}^{vac} $, although it might be sensible to do so, given its large
radiative (quantum) content \cite{ven90}.
We note here that the former is expected to contain the fundamental 
length scale $\xi$ as well, in the form $ \simeq + (3 / \xi^2) \, g_{\mu\nu} $.}
\beq
T_{\mu\nu}^{vac} \, \equiv \,  {\delta G(\Box) \over G_0} \, T_{\mu\nu} \; .
\label{eq:tmunu_vac1}
\eeq
In addition, consistency of the full nonlocal field equations requires that the sum
be conserved,
\beq
\nabla^\mu \left ( T_{\mu\nu}   +  T_{\mu\nu}^{vac}  \right ) = 0 \; .
\eeq
It is important to note at this stage that the nature of the covariant d'Alembertian  
$ \Box \equiv g^{\mu\nu} \, \nabla_\mu \nabla_\nu  $ is such that the result
depends on the type of the object it acts on. 
Here $T_{\mu\nu} $ is a second rank tensor (as in Eq.~(\ref{eq:box_on_tensors})), 
which causes a re-shuffling of components in $T_{\mu\nu} $ due to the matrix nature
of both tensor $\Box$ and tensor $G(\Box)$, and eventually accounts for the 
generation of a non-vanishing induced pressure term.
This is clearly seen in the effective field equations of Eqs.~(\ref{eq:fried_tt})
and (\ref{eq:fried_rr}), and in the ensuing definitions of Eq.~(\ref{eq:rhovac_t}).

In general though one cannot expect that the contribution $ T_{\mu\nu}^{vac} $
will always be expressible in the perfect fluid form of Eq.~(\ref{eq:tmunu_perf}), even if the
original $ T_{\mu\nu} $ for matter (or radiation) has such a form.
The former will in general contain, for example, non-vanishing shear stress contributions, 
even if they were originally absent in the matter part.
Nevertheless the interesting question arises of whether, for example, $w_{vac}= \third $
continues to hold beyond the FLRW case treated above.
In part this question will be answered affirmatively below, in the case of matter density perturbations.

\newpage

\vskip 40pt
\newsection{Relativistic Treatment of Matter Density Perturbations}
\hspace*{\parindent}
\label{sec:pert}


Besides the modified cosmic scale factor evolution just discussed, 
the running of $G(\Box)$ given in Eq.~(\ref{eq:grun_box})
also affects the nature of matter density perturbations on very large scales.
In computing these effects, it is customary to introduce a perturbed metric of
the form
\beq
{d\tau}^2 = {dt}^2 - a^2 \left ( \delta_{ij} + h_{ij} \right ) dx^i dx^j \; ,
\label{eq:pert_metric}
\eeq
with $a(t)$ the unperturbed scale factor and $ h_{ij} ({\bf x},t)$ a small
metric perturbation, and $h_{00}=h_{i0}=0$ by choice of coordinates.
As will become clear later, we will mostly be concerned here with the trace
mode $h_{ii} \equiv h $, which determines the nature of matter density
perturbations.
After decomposing the matter fields into background and fluctuation contribution, 
$\rho = \bar{\rho}+\delta \rho$, $p = \bar{p}+\delta p $, and ${\bf v} = \bar{\bf v}+\delta {\bf v}$, 
it is customary in these treatments to expand the density, pressure and metric trace 
perturbation modes in spatial Fourier modes,
\bea
\delta \rho ({\bf x},t) & = & \delta \rho_{\bf q} (t) \, e^{i \, {\bf q} \, \cdot \, {\bf x}}
\;\;\;\;\;\;\;\;
\delta p ({\bf x},t) = \delta p_{\bf q} (t) \, e^{i \, {\bf q} \,\cdot \, {\bf x}}
\nonumber \\
\delta {\bf v} ({\bf x},t) & = & {\delta {\bf v}}_{\bf q} (t)  \, e^{i \, {\bf q} \, \cdot \, {\bf x}}
\;\;\;\;\;\;\;\;
h_{ij} ({\bf x},t) = h_{ {\bf q} \, ij} (t)\, e^{i \, {\bf q} \, \cdot \, {\bf x}} 
\eea
with ${\bf q}$ the comoving wavenumber.
Once the Fourier coefficients have been determined, the original perturbations can
later be obtained from
\beq
\delta \rho ({\bf x},t) \, = \, \int { d^3 {\bf x} \over ( 2 \pi )^{3/2} }
\, e^{ - i \, {\bf q} \, \cdot \, {\bf x}} \, \delta \rho_{\bf q} (t) 
\eeq
and similarly for the other fluctuation components.
Then the field equations with a constant $G_0$ (Eq.~(\ref{eq:field}))
are given to zeroth order in the perturbations by
Eq.~(\ref{eq:fried_0}),  which fixes the three background fields 
$a(t)$, $\bar{\rho} (t)$ and $\bar{p} (t) $.
Note that in a comoving frame the four-velocity appearing in Eq.~(\ref{eq:tmunu_perf})
has components $ u^i = 1, \; u^0 = 0 $.
To first order in the perturbations and in the limit ${\bf q} \rightarrow 0$ the field equations give
\bea
{\dot{a} (t) \over a (t)}\, \dot{h} (t) & = & 8 \pi \, G_0  \, \bar{\rho} (t) \, \delta (t) 
\nonumber \\
\ddot{h} (t) + 3 \, {\dot{a} (t) \over a (t)}\, \dot{h} (t) 
& = & - \, 24 \pi \, G_0  \, w \, \bar{\rho}(t) \, \delta (t)
\eea
with the matter density contrast defined as $\delta (t) \equiv \delta \rho (t) / \bar{\rho} (t) $, 
$h(t) \equiv h_{ii} (t)$ the trace part of $h_{ij}$, and $w=0$ for non-relativistic matter.
When combined together, these last two equations then yield 
a single equation for the trace of the metric perturbation,
\beq
\ddot{h} (t) + 2 \, {\dot{a} (t) \over a (t)} \, \dot{h} (t) \; = \; 
 - \, 8 \pi \, G_0 ( 1 + 3\, w ) \, \bar{\rho}(t) \, \delta (t) \; .
\eeq
From first order energy conservation one has 
$ - {1 \over 2}\, \left ( 1 + w \right )\, h (t) = \delta (t) $, 
which then allows one to eliminate $h(t)$ in favor of $\delta(t)$.
This finally gives a single second order equation for the density contrast $\delta(t)$,
\beq
\ddot{\delta} (t) + 2 \, {\dot{a} \over a} \, \dot{\delta} (t) 
- 4 \pi \, G \, \bar{\rho}(t) \, \delta(t) = 0 \; .
\eeq
In the case of a running $G(\Box)$ these equations need to be re-derived
from the effective covariant field equations of Eq.~(\ref{eq:field1}), and lead to several
additional terms not present at the classical level.
Not surprisingly, as we shall see below, the correct field equations with a running $G$ 
are not given simply by a naive replacement $G \rightarrow G(t) $, which would
lead to incorrect results, and violate general covariance.

\vskip 40pt
\subsection{Zeroth Order Energy-Momentum Conservation}
\hspace*{\parindent}
\label{sec:enmom_zeroth}

As a first step in computing the effects of density matter perturbations one
needs to examine the consequences of energy and momentum conservation,
to zeroth and first order in the relevant perturbations.
If one takes the covariant divergence of the field equations in Eq.~(\ref{eq:field1}),
the left hand side has to vanish identically because of the Bianchi identity. 
The right hand side then gives 
$\nabla^\mu \left (  T_{\mu\nu} + T_{\mu\nu}^{vac} \right ) =0 $,
where the fields in $ T_{\mu \nu}^{vac} $ can be expressed, at least to lowest order,
in terms of the $p_{vac}$ and $\rho_{vac}$ fields defined in Eqs.~(\ref{eq:rhovac_t}) and (\ref{eq:wvac_def}).
The first equation one obtains is the zeroth (in the fluctuations) order energy conservation
in the presence of $G(\Box)$, which reads
\beq
3 \, {\dot{a} (t) \over a (t)} \, 
\left [ \left (1+w \right ) + \left (1+ w_{vac} \right )\,{\delta G(t) \over G_0} 
\right ]  \bar{\rho} (t) 
 +   { \dot{\delta G}(t) \over G_0} \, \bar{\rho} (t)
 + \left ( 1 + {\delta G(t) \over G_0} \right )\, \dot{\bar{\rho}} (t)  = 0 \; .
\label{eq:encons_zeroth_w}
\eeq
For $w=0$ and $w_{vac}={1 \over 3}$ this reduces to
\beq
\left [ 3 \, {\dot{a} (t) \over a (t)} 
+ 4 \, {\dot{a} (t) \over a (t)}\,{\delta G(t) \over G_0} 
+  { \dot{\delta G}(t) \over G_0} \right ] \, \bar{\rho}(t) 
+ \left ( 1 + {\delta G(t) \over G_0} \right ) \, \dot{\bar{\rho}} (t) = 0 \; ,
\label{eq:encons_zeroth}
\eeq
or equivalently in terms of the variable $a(t)$ only
\beq
\left [
{ 3 \over a } + { 4 \over a } { \delta G (a) \over G_0 } + { \delta G' (a) \over G_0 }
\right ] 
\bar{\rho} (a) 
+ \left (  1  + { \delta G (a) \over G_0 } \right ) \bar{\rho}' (a) = 0 \; .
\eeq
In the absence of a running $G$ these equations reduce to the ordinary mass
conservation equation for $w=0$,
\beq
\dot{\bar{\rho}}(t) = - 3 \, { \dot{a}(t) \over a(t)} \, \bar{\rho}(t) \; .
\label{eq:encons_frw}
\eeq
It will be convenient in the following to solve the energy conservation
equation not for ${\bar{\rho}} (t)$, but instead for ${\bar{\rho}} (a)$.
This requires that, instead of using the expression for $G(t)$ in Eq.~(\ref{eq:grun_t}),
one uses the equivalent expression for $G(a)$
\beq
G (a) = G_0 \left ( 1 + {\delta G (a) \over G_0}  \right ), 
\;\;\;\;\;\; {\rm with}
\;\; {\delta G(a) \over G_0} \equiv  c_a \, \left ( {a \over a_0 } \right )^{\gamma_\nu}  + \dots \; .
\label{eq:grun_a}
\eeq
In this last expression the power is $ \gamma_\nu = 3/2 \nu $, since from Eq.~(\ref{eq:grun_t}) 
one has for non-relativistic matter $a(t)/a_0 \approx (t/t_0)^{2/3}$ in the absence of a running $G$.
In the following we will almost exclusively consider the case $\nu= \third$ \cite{ham00} for which
therefore $\gamma_\nu = 9/2$.
\footnote{This implicitly assumes that the cosmic evolution is largely matter dominated,
if $p=w \rho$ then $ a(t)/a_0 = (t / t_0)^{ 2 / 3 (1+w) }$.
In the opposite regime where a cosmological constant can eventually prevail one has
instead $ a(t)/a_0 =\exp \sqrt{ \lambda / 3} (t - t_0)$. 
Then $ {t \over t_0 } = 1 + {1 \over t_0 } \sqrt{ 3 \over \lambda} \log { a \over a_0 }$
and for $ t_0 \simeq \xi $ and $\sqrt{ 3 \over \lambda } \simeq \xi $ one has simply
$ { t \over t_0 } = 1 + \log { a \over a_0 } $.}
Then in the above expression $c_a \approx c_t$ if $a_0$ is identified with a scale factor 
appropriate for a universe of size $\xi$; to a good approximation this should
correspond to the universe ``today'',  with the relative scale factor customarily normalized
at such a time to $a/a_0=1$.
Consequently, and with the above proviso, the constant $c_a$ in Eq.~(\ref{eq:grun_a})
can safely be taken to be of the 
same order as the constant $c_0$  appearing in the original expressions for $G(\Box)$ in
Eq.~(\ref{eq:grun_box}).

Then the solution to Eq.~(\ref{eq:encons_zeroth}) can be written as
\beq
\bar{\rho} (a)= {\rm const. } \; \exp \left \{
- \int  { d a \over a} \;
\left (  3 + { \delta G (a) \over G_0 } + a \, { \delta G' (a) \over G_0  } \right ) 
\right \} \; ,
\label{eq:rho_zeroth_sim}
\eeq
or, more explicitly, as
\beq
\bar{\rho} (a)= \bar{\rho}_0 \, \left ( {a_0 \over a} \right )^3 \,   
\left ( 
{ 1 + c_a \over  1 + c_a \, \left ( {a \over a_0} \right )^{\gamma_\nu} } 
\right )^{ (1+ \gamma_\nu) / \gamma_\nu}
\simeq \; 
\bar{\rho}_0 \, \left ( {a_0 \over a} \right )^3 \,   
{ 1 + ( 1 + \gamma_\nu^{-1} ) \, c_a \over  
1 + ( 1 + \gamma_\nu^{-1} )  \, c_a \, \left ( {a \over a_0} \right )^{\gamma_\nu} } 
\label{eq:rho_zeroth}
\eeq
with $\bar{\rho} (a) $ normalized so that $ \bar{\rho} (a=a_0) = \bar{\rho}_0$.
For $c_a=0$ the above expression reduces of course to the usual result for 
non-relativistic matter, 
\beq
\bar{\rho}(t) = \bar{\rho}_0 \, \left ( {a_0 \over a} \right ) ^3 \; .
\label{eq:rho_fried}
\eeq
Furthermore, here one also finds that the zeroth order momentum conservation 
equation is identically satisfied, just as in the case of constant $G$.

\vskip 40pt
\subsection{Zeroth Order Field Equations with Running ${\bf G(\Box)}$}
\hspace*{\parindent}
\label{sec:field_zeroth}

The zeroth order field equations with the running of $G$ included 
were already given in Eq.~(\ref{eq:fried_run}).
One can subtract the two equations from each other to get
an equation that does not contain $\lambda$,
\beq
{{\dot{a}}^2 (t) \over {a}^2 (t)}  - {\ddot{a} (t) \over a (t)} = 
4 \pi G_0 \left [ \left (1+w \right ) 
+ \left ( 1+w_{vac} \right )\,{\delta G(t) \over G_0} \right ] \bar{\rho}(t) \; .
\label{eq:fried_run_nolambda}
\eeq
Alternatively, from Eqs.~(\ref{eq:fried_run}) one can obtain a single 
equation that only involves the acceleration term with $ \ddot{a} (t)$,
\beq
3 \, {\ddot{a} (t) \over a (t)} = - 4 \pi G_0 
\left [ 
\left ( 1 + 3\, w \right ) 
+ \left ( 1 + w_{vac} \right )\, {\delta G(t) \over G_0} 
\right ] \bar{\rho} (t) + \lambda \; .
\label{eq:fried_run_dotdota}
\eeq
It is also rather easy to check the overall consistency of the energy conservation equation, 
Eq.~(\ref{eq:encons_zeroth}),  and of the two field equations in Eq.~(\ref{eq:fried_run}).
This is done by (i) taking the time derivative of the first $ tt $ equation in Eq.~(\ref{eq:fried_run}),
(ii) replacing terms involving $ \dot{\bar{\rho}} $ by ${\bar{\rho}}$ using the energy conservation 
equation, Eq.~(\ref{eq:encons_zeroth}), and (iii) finally by substituting again the result of
the first ($ tt $) equation into Eq.~(\ref{eq:fried_run}) to obtain the second ($rr$) equation in Eq.~(\ref{eq:fried_run}).

\vskip 40pt
\subsection{Effective Energy-Momentum Tensor ${\bf \rho_{vac}}$, ${\bf p_{vac}}$}
\hspace*{\parindent}
\label{sec:enmom_vac}

The next step consists in obtaining the equations which govern the effects of 
small field perturbations.
These equations will involve, apart from the metric perturbation $h_{ij}$, the matter 
and vacuum polarization contributions.
The latter arise from (see Eq.~(\ref{eq:tmunu_vac}))
\beq
\left ( 1 + {\delta G(\Box) \over G_0}  \right ) \, T_{\mu\nu}  \, = \,
T_{\mu\nu}   +  T_{\mu\nu}^{vac} 
\eeq
with a nonlocal 
$ T_{\mu\nu}^{vac} \equiv ( \delta G(\Box) / G_0 ) \, T_{\mu\nu} $.
Fortunately to zeroth order in the fluctuations the results of Ref. \cite{hw05} indicated 
that the modifications from the nonlocal vacuum polarization term could 
simply be accounted for by the substitution
\beq
\bar{\rho} (t) \rightarrow \; \bar{\rho} (t) + {\bar\rho}_{vac} (t)     
\;\;\;\;\;\;\;\;\;\;\;\;
\bar{p} (t) \rightarrow \; \bar{p} (t) +  {\bar p}_{vac} (t) \; .
\label{eq:sub0}
\eeq
Here we will apply this last result to the small field fluctuations as well, and set
\beq 
\delta \rho_{\bf q} (t) \rightarrow \; \delta \rho_{\bf q} (t) + \delta \rho_{{\bf q} \, vac} (t)
\;\;\;\;\;\;\;\;\;\;\;\;
\delta p_{\bf q}  (t) \rightarrow \; \delta p_{\bf q}  (t) +  \delta p_{{\bf q} \, vac} (t) \; .
\label{eq:sub1}
\eeq
The underlying assumptions is of course that the equation of state for the vacuum fluid still
remains roughly correct when a small perturbation is added.
Furthermore, just like we had $ {\bar p} (t) = w \,  \bar{\rho} (t) $ (Eq.~(\ref{eq:w_def}))
and  $\bar{p}_{vac} (t) = w_{vac} \,  \bar{\rho}_{vac}(t) $ (Eq. ~(\ref{eq:wvac_def}))
with $ w_{vac} = \third $, we now write for the fluctuations
\beq
\delta p_{\bf q}  (t) = w \, \delta \rho_{\bf q}   (t) \;\;\;\;\;\;\;\;\; 
\delta p_{{\bf q} \, vac} (t) = w_{vac} \, \delta \rho_{{\bf q} \, vac} (t) \; ,
\label{eq:wvac_fluc}
\eeq
at least to leading order in the long wavelength limit, ${\bf q} \rightarrow 0 $.
In this limit we then have simply
\beq
\delta p (t) = w \, \delta \rho   (t) \;\;\;\;\;\;\;\;\; 
\delta p_{vac} (t) = w_{vac} \, \delta \rho_{vac} (t) \equiv 
w_{vac}  \, {\delta G(t) \over G_0}  \delta \rho (t) \; ,
\label{eq:wvac_fluc1}
\eeq
with $G(t)$ given in Eq.~(\ref{eq:grun_t}), and we have used Eq.~(\ref{eq:rhovac_t}), 
now applied to the fluctuation $\delta \rho_{vac} (t)$,
\beq
\delta \rho_{vac} (t) \, = \, {\delta G(t) \over G_0}  \, \delta \rho (t) + \dots
\label{eq:delta_rhovac_t}
\eeq
where the dots indicate possible additional $O(h)$ contributions.

A bit of thought reveals that the above treatment is incomplete,
since $G(\Box)$ in the effective field equation of Eq.~(\ref{eq:field1}) 
contains, for the perturbed RW metric of Eq.~(\ref{eq:pert_metric}), 
terms of order $h_{ij}$,  which need to be accounted for in the effective $T^{\mu\nu}_{vac}$.
Consequently the covariant d'Alembertian has to be Taylor expanded in the small
field perturbation $h_{ij}$,
\beq
\Box (g) \, = \, \Box^{(0)}  + \Box^{(1)} (h) + O (h^2) \; ,
\eeq
and similarly for $G(\Box)$
\beq
G(\Box) \, = \, G_0 \,  \left [ 
1 + \, { c_0 \over \xi^{1 / \nu} } \,  
\left ( { 1 \over \Box^{(0)} + \Box^{(1)} (h) + O(h^2) } \right )^{1/ 2 \nu}  + \dots
\right ]  \; ,
\label{eq:gbox_h}
\eeq
which requires the use of the binomial expansion for the operator
$ (A+B)^{-1} = A^{-1} - A^{-1} B A^{-1}  + \dots $.
Thus for sufficiently small perturbations it should be adequate to expand $G(\Box)$ 
entering the effective field equations in powers of the metric perturbation $h_{ij} $.
Since a number of subtleties arise in this expansion, we shall first consider
the simpler case of a scalar box, where some of the issues we think can be
clearly identified, and addressed.
After that, we will consider the more complicated case of the tensor box.
This will be followed by a determination of the effects of the running of $G$ on the relevant
matter and metric perturbations, by the use of the modified field equations,
now expanded to first order in the perturbations.

\vskip 40pt
\subsection{${\bf O(h)}$ Correction using Scalar Box}
\hspace*{\parindent}
\label{sec:s_box}

In this section the term $O(h)$ in $\delta \rho_{vac}$ of Eq.~(\ref{eq:sub1}) will be 
determined, using a set of formal manipulations involving the covariant scalar box operator.
Instead of considering the full field equations with a running $ G (\Box) $, 
as given in Eq.~(\ref{eq:field1}),
\beq
R_{\mu \nu} - {1 \over 2} \, g_{\mu \nu} \, R  + \lambda \, g_{\mu \nu}= 8 \pi \, G_0 \, 
\left ( 1 + {\delta G(\Box) \over G_0} \right ) \, T_{\mu \nu}
\label{eq:field2}
\eeq
we will consider here instead the action of a scalar $ G (\Box) $ on the trace of 
the field equations for $\lambda=0$,
\beq
R = - 8 \pi G_0 \, \left ( 1 + {\delta G(\Box) \over G_0} \right )  T_{\lambda}^{\;\; \lambda} \; ,
\label{eq:field_trace}
\eeq
or equivalently, by having the operator $ G (\Box) $ act on the left hand side,
\beq
\left ( 1 - {\delta G(\Box) \over G_0} + \dots \right ) R 
= - 8 \pi \, G_0 \, T_{\lambda}^{\;\; \lambda} \; .
\eeq
For a perfect fluid one has simply $ T_{\lambda}^{\;\; \lambda} = - \rho $,  which then gives \cite{hw05}
\beq
G_0 \, \left ( 1 + {\delta G(\Box) \over G_0} \right )  T_{\lambda}^{\;\; \lambda}
\;\; \rightarrow \;\;
G_0 \, \left [ 1 + c_t \, \left ( { t \over t_0} \right )^{1 / \nu} + \dots \right ] \, T_{\lambda}^{\;\; \lambda}
\equiv G(t) \, T_{\lambda}^{\;\; \lambda} \; ,
\label{eq:gbox_scal}
\eeq
or equivalently
\beq
G_0 \, \left [ 1 + c_t \, \left ( { t \over t_0} \right )^{1 / \nu} + \dots \right ] \, \bar{\rho} (t)
\equiv  G(t) \, \bar{\rho} (t) \; ,
\label{eq:gbox_scal_rho}
\eeq
with $c_t \simeq 0.785 \, c_0 $, and $t_0 = \xi$ \cite{hw05} (in the tensor box case a slightly
smaller value was found, $c_t \simeq 0.450 \, c_0 $).
The two terms in Eq.~(\ref{eq:gbox_scal_rho}) are of course recognized, up to a factor
of $G_0$, as the combination
\beq
\bar{\rho} (t) + {\bar\rho}_{vac} (t)     
\eeq
of Eq.~(\ref{eq:sub0}), with 
${\bar\rho}_{vac} (t) \equiv \delta G (t) / G_0 \cdot \bar{\rho} (t) $.
Thus the zeroth order result obtained by the use of the scalar 
d'Alembertian acting on the trace of the field equations is consistent with what has been
used so far for $G(t)$.

To compute the higher order terms in the $h_{ij}$'s appearing in the metric of
Eq.~(\ref{eq:pert_metric}) one needs to expand $G(\Box)$ according to 
Eq.~(\ref{eq:gbox_h}) giving
\beq
G(\Box) = G_0 \, \left [ 
1 + \, { c_0 \over \xi^{1 / \nu} } \, 
\left (
\left ( { 1 \over \Box^{(0)} } \right )^{1 / 2 \nu} 
- {1 \over 2 \, \nu} \, { 1 \over \Box^{(0)} } \cdot \Box^{(1)} (h)  \cdot
\left ( { 1 \over \Box^{(0)} } \right )^{1 / 2 \nu} \, + \dots
\right )
\right ] \; .
\label{gbox_h_expanded}
\eeq
Here we are interested in the correction of order $h_{ij}$, when the above operator
acts on the scalar $ T_{\lambda}^{\;\; \lambda} = - {\bar \rho} $.
This would then give the correction $O(h)$ to $\delta \rho_{vac} $, namely the second term in
\beq
\delta \rho_{vac} (t) = { \delta G(\Box^{(0)}) \over G_0 }\; \delta \rho (t)
+ {\delta G (\Box) (h) \over G_0} \;  \bar{\rho} (t) \; ,
\label{eq:delta_rhovac_h}
\eeq
with the first term being simply given in the FLRW background
by $\delta G(t)/G_0 \cdot \delta \rho (t) $.
Here the $O(h)$  correction is given explicitly by the expression
\beq
{\delta G(\Box) (h) \over G_0} \, {\bar \rho} \, =  \, 
- {1 \over 2 \, \nu}  \, { c_0 \over \xi^{{1 / \nu}} }  \,
{ 1 \over \Box^{(0)} } \cdot \Box^{(1)} (h)   \cdot
\left ( { 1 \over \Box^{(0)} } \right )^{ 1/ 2\nu}   \cdot
{\bar \rho} \; .
\eeq
The effect of the $ ( \Box^{(0)} )^{ -1/ 2\nu} $ term is essentially to make the
coupling time dependent, {\it i.e.} to correctly reproduce the required overall 
time-dependent factor $\delta G(t)/G_0$.

Now the scalar d'Alembertian $\Box = g^{\mu\nu} \, \nabla_\mu \nabla_\nu $
acting on scalar functions $S(x)$ has the form
\beq
\Box \, S(x) \equiv {1 \over \sqrt{g} } \, 
\partial_\mu \, g^{\mu\nu} \sqrt{g} \, \partial_\nu \, S(x)
\eeq
In the absence of $h_{ij}$ fluctuations this gives for the metric in Eq.~(\ref{eq:pert_metric})
\beq
\Box^{(0)} S(x) = 
{1 \over a^2} \nabla ^2 S - 3\, {\dot{a} \over a} \, \dot{S} - \ddot{S}
\;\; \rightarrow \;\; 
\left ( - \partial_t^2 - 3 \, {\dot{a} \over a} \partial_t  \right ) S(t) \; ,
\eeq
where in the second expression we have used the properties of the RW 
background metric: we only need to consider
functions that are time dependent,  so that $ S ({\bf x}, t) \rightarrow S (t) $.
To first order in the field fluctuation $h_{ij}$ of Eq.~(\ref{eq:pert_metric}) one computes
\beq
\Box^{(1)} (h) \, S(x) =  - { 1 \over 2} \dot{h} \, \dot{S} 
- {1 \over a^2}  h_{xx} \, \partial_x^2 S 
+ {1 \over a^2} \left ( - \partial_x \, h_{xx} \right ) \cdot \partial_x S 
+ {1 \over 2\, a^2} \, \partial_x h \cdot \partial_x S + \dots
 \eeq
with the trace $ h(t) = h_{xx}(t) + h_{yy}(t) + h_{zz}(t) $.
But for a function of the time only one obtains
\beq
\Box^{(1)} (h) \, \rho (t) = - { 1 \over 2} \, \dot{h} (t) \, \dot{S} (t) \; .
\eeq
Thus to first order in the fluctuations one obtains the expression
\beq
{ 1 \over \Box^{(0)} } \cdot \Box^{(1)}  (h) \cdot \left ( \, \delta G \,  \bar{\rho} \, \right )=
{ 1 \over - \partial_t^2 - 3 \, {\dot{a} \over a} \, \partial_t } \cdot
\half \, \dot{h} \left ( 3 \, {\dot{a} \over a} \, \delta G - \dot{\delta G} \right ) \bar{\rho}
\label{eq:box_ratio}
\eeq
where use has been made of the zeroth order mass conservation equation
in Eq.~(\ref{eq:encons_frw}).
Note that this result also correctly incorporates the effect of $G (\Box^{(0)})$ on functions of $t$, 
as given for example in Eq.~(\ref{eq:gbox_scal}), which ensures the proper 
running of $\delta G(t)$.

Now in our treatment we are generally interested in mass density and metric perturbations 
around a near-static background described by $ \dot{a} / a = H(t) $, and $\bar{\rho} (t)$.
For these we expect the relevant time variations in $\delta \rho$ and $h$ to be somewhat 
larger than for the background itself.
Thus for sufficiently slowly varying background fields we retain only $h(t)$ and its derivatives, 
and for a sufficiently slowly varying $h(t)$ only $h(t)$ and the lowest derivatives.
Then the factors of $ \dot{a} / a $ are seen to cancel out at leading order between 
numerator and denominator in Eq.~(\ref{eq:box_ratio}), and one is left simply with 
\beq
{ 1 \over \Box^{(0)} } \cdot \Box^{(1)}  (h) \cdot \delta G (t) \, \bar{\rho} (t) \,
= \, - \, { 1 \over 2}  \, \delta G (t) \, h(t) \, \bar{\rho} (t) + \dots 
\eeq
Putting everything together, one finds for the $O(h)$ correction
\beq
 {\delta G(\Box) (h) \over G_0} \; \bar{\rho} (t)  
\simeq + \, {1 \over 4 \, \nu} \,  \, {\delta G(t) \over G_0}  \, h(t) \, \bar{\rho} (t) \; .
\eeq
The scalar box calculation just described allows one to compute the correction $O(h)$
to $\delta \rho_{vac} (t) $  in Eq.~(\ref{eq:delta_rhovac_h}), and leads to the following 
$O(h)$ modification of Eq.~(\ref{eq:delta_rhovac_t})
\beq
\delta \rho_{vac} (t)  =  {\delta G(t) \over G_0}\, \delta \rho (t) + 
{1 \over 2 \, \nu} \, c_h \, {\delta G(t) \over G_0}  \; h(t) \, \bar{\rho}(t)
\label{eq:delta_rhovac_t_h}
\eeq
and similarly from $ \delta p_{vac} (t) =  w_{vac}  \delta \rho_{vac} (t)  $,
\beq
\delta p_{vac} (t) =  w_{vac}  \, \left ( {\delta G(t) \over G_0} \, \delta \rho (t) 
+ {1 \over 2 \, \nu} \, c_h \, {\delta G(t) \over G_0} \,  h (t) \, \bar{\rho} (t) \right )
\label{eq:delta_pvac_t_h}
\eeq
with $w_{vac}=\third$.
The second $O(h)$ terms in both expressions account for the feedback of the 
metric fluctuations $h$ on the 
vaccum density $\delta \rho_{vac}$ and pressure $\delta p_{vac}$ fluctuations.

The potential flaw with the preceding argument is that it assumes that certain 
very specific functions of the background stay constant, or at least very slowly varying.
In the case at hand this was $ \dot{a} / a \equiv H(a) \approx {\rm const.} $ and 
$\rho \approx {\rm const.}$, which in principle is not the only possibility,
and would seem therefore a bit restrictive.
A slightly more general approach, and a check, to the evaluation of the expression in
Eq.~(\ref{eq:box_ratio}) goes as follows.
One assumes instead a harmonic time dependence for the
metric fluctuation $h(t) = h_0 \, e^{i \omega t}$, and similarly for 
$a(t)=a_0 \, e^{i \Gamma t}$, ${\bar \rho} (t)= {\bar \rho_0} \, e^{i \Gamma t}$,
and $\delta G (t)= \delta G_0 \, e^{i \Gamma t}$;
different frequencies for $a$ and ${\bar \rho}$ could be considered as well, but here 
we will just stick with the simplest possibility.
Then from the last expression in Eq.~(\ref{eq:box_ratio}) one has
\beq
{ 1 \over - \partial_t^2 - 3 \, {\dot{a} \over a} \, \partial_t } \cdot
\half \, \dot{h} 
\left ( 3 \, {\dot{a} \over a} \, \delta G - \dot{\delta G} \right ) \bar{\rho}
\, = \, { 1 \over \omega^2 + 7 \Gamma \omega + 10 \Gamma^2 } \cdot
\left ( - \, \Gamma \, \omega \, \delta G \, h \, \bar{\rho}  \right ) \; .
\label{eq:box_ratio_omega}
\eeq
In the limit $\omega \gg \Gamma$, corresponding to $ \dot{h}  / h  \gg  \dot{a} / a  $, 
one obtains for the above expression
\beq
- \, { \Gamma \over \omega } \, \delta G (t) \, h (t) \, \bar{\rho} (t) 
\, \simeq \, - \, \left ( {\dot{a} \over a}  \, {h \over \dot{h} } \right ) 
\, \delta G (t) \, h (t) \, \bar{\rho} (t) \; ,
\eeq
after substituting back $ \dot{h} / h= i \omega$ and $\dot{a}/ a= i \Gamma$ in the last
expression.
Then $ \delta \rho_{vac} (t)  $  in Eq.~(\ref{eq:delta_rhovac_t_h}) 
now involves the quantity $c_h$
\beq
c_h \; = \; {\dot{a} \over a}  \, { h \over \dot{h} } \; .
\eeq
At first this last factor (a function and not a constant) would seem rather hard to
evaluate, and perhaps not even close to constant in time.
But a bit of thought reveals that, to the order we are working, one can write
\beq
{ \dot{h}  \over h} \, {a \over \dot{a} } 
\; = \; { \partial \log h (a) \over \partial \log a }
\; = \; { \partial \log \delta (a)  \over \partial \log a } 
\equiv  f(a) \; ,
\label{eq:ch_fa}
\eeq
where $\delta (a)$ is the matter density contrast, and 
$f(a)$ the known density growth index \cite{pee93}.
In the absence of a running $G$ (which is all that is needed, to the order
one is working here) an explicit form for $f(a)$ is known in terms of
derivatives of a Gauss hypergeometric function, which will be given below.
One can then either include the explicit form for $f(a)$ in the above formula for
$ \delta \rho_{vac} (t) $, or use the fact that for a scale factor referring
to ``today'' $a/a_0 \approx 1$, and for a matter fraction $\Omega \approx 0.25$,
one knows that $ f (a=a_0) \simeq 0.4625 $, and thus in Eq.~(\ref{eq:delta_rhovac_t_h})
one obtains the improved result $c_h \simeq 2.1621 $.
This can then be compared to the earlier result, which gave $c_h \simeq 1/2 $.

A similar analysis can now be done in the opposite, but in our opinion less physical, 
$\omega \ll \Gamma$ limit,
for which one now obtains for the expression in Eq.~(\ref{eq:box_ratio_omega})
\beq
- {1 \over 10 } \, 
\left ( {a \over \dot{a} }  \, {\dot{h} \over h } \right )
\, \delta G (t) \, h (t) \, \bar{\rho} (t) \; .
\eeq
This new limit is less physical because of the fact 
that now the background is assumed to be varying more rapidly in time 
than the metric perturbation itself,  $ \dot{a} / a  \gg \dot{h}  / h $.
For $ \delta \rho_{vac} (t)  $ one then obtains a similar expression to
the one in Eq.~(\ref{eq:delta_rhovac_t_h}),  with a different coefficient
\beq
c_h \; = \; {1 \over 10 } \, \, {a \over \dot{a} }  \, {\dot{h} \over h }
\eeq
still involving the quantity $ ( a / \dot{a} ) ( \dot{h} / h ) \equiv f(a)$.
By the same chain of arguments used in the previous paragraph one can now 
either include the explicit form for $f(a)$ in the formula for $ \delta \rho_{vac} (t) $, 
or use the fact that for a scale factor referring
to ``today'' $a/a_0 \approx 1$ and a matter fraction $\Omega \approx 0.25$
one knows that $ f (a=a_0) \simeq 0.4625 $.
In this case one then has in Eq.~(\ref{eq:delta_rhovac_t_h}) 
$c_h \simeq (1/10) \times 0.4625 = 0.0463 $.
One disturbing, but not entirely surprising, general aspect of the whole calculation in
this second $\omega \ll \Gamma$ limit (as opposed to the previous treatment 
in the opposite limit) is its rather significant sensitivity, in the final result, to the set of 
assumptions initially made about the time development of the background
as specified by the functions $a(t)$ and $\bar{\rho} (t)$.
Therefore in the following we shall not consider this limit further.

To summarize, the results for a scalar box and a slowly varying background,
$ \dot{h}  / h  \gg  \dot{a} / a  $, give the $O(h)$ corrected expression
for $\delta \rho_{vac} (t)$ in Eq.~(\ref{eq:delta_rhovac_t_h}) and 
$\delta p_{vac} (t) = w_{vac} \, \delta \rho_{vac} (t) $,  with $c_h \simeq + 2.1621 $.

\vskip 40pt
\subsection{${\bf O(h)}$ Correction using Tensor Box}
\hspace*{\parindent}
\label{sec:t_box}

The results of Eqs.~(\ref{eq:delta_rhovac_h}) and (\ref{eq:delta_rhovac_t_h}) for
the vacuum polarization contribution,
\beq
\delta \rho_{vac} (t)  =  {\delta G(t) \over G_0}\, \delta \rho (t) + 
{1 \over 2 \, \nu} \, c_h \, {\delta G(t) \over G_0}  \; h(t) \, \bar{\rho}(t)
\eeq
and similarly for $ \delta p_{vac} (t) =  w_{vac} \, \delta \rho_{vac} (t)  $ with $w_{vac}= \third$,
were obtained using a scalar d'Alembertian to implement $G(\Box)$ by considering the trace
of the field equation, Eq.~(\ref{eq:field_trace}).
In this section we will discuss instead the result for the full tensor d'Alembertian, 
as it appears originally in the effective field equations of 
Eqs.~(\ref{eq:field1}) and (\ref{eq:field2}).

Now the d'Alembertian operator $ \Box \; = \; g^{\mu\nu} \nabla_\mu \nabla_\nu $
acts on the second rank tensor $T_{\mu\nu}$ as in Eq.~(\ref{eq:box_on_tensors}),
and should therefore be regarded as a four by four matrix,
transforming $T_{\mu\nu} $ into $[ \Box T ]_{\mu\nu}$.
Indeed it is precisely this matrix nature of $\Box$, and therefore of $G(\Box)$,
that accounts for the fact that a vacuum pressure is induced in the first place,
leading to a $w_{vac} \neq 0 $.

To compute the correction of $O(h)$ to $ \delta \rho_{vac} (t) $ one needs to consider
the relevant term in the expansion of $ ( 1 + \delta G(\Box) / G_0 ) \, T_{\mu \nu} $,
which we write as
\beq
- {1 \over 2 \, \nu} \, { 1 \over \Box^{(0)} } \cdot \Box^{(1)} (h) \cdot
{ \delta G ( \Box^{(0)} ) \over G_0 } \cdot T_{\mu \nu}
\; = \; 
- {1 \over 2 \, \nu} \, { c_0 \over \xi^{1 / \nu} } \,
{ 1 \over \Box^{(0)} } \cdot \Box^{(1)} (h)  \cdot 
\left ( { 1 \over \Box^{(0)} } \right )^{1 / 2 \nu} \cdot T_{\mu \nu} \; .
\label{gbox_h_tensor}
\eeq
This last form allows us to use the results obtained previously 
for the FLRW case in \cite{hw05}, namely
\beq
{ \delta G ( \Box^{(0)} ) \over G_0 } \, T_{\mu \nu} \; = \; T_{\mu \nu}^{vac}
\eeq
with here
\beq
T_{\mu \nu}^{vac} \; = \;
\left [ p_{vac} (t) + \rho_{vac} (t) \right ] u_\mu \, u_\nu + g_{\mu \nu} \, p_{vac} (t)
\label{eq:tmunu_vac_2}
\eeq
and (see Eqs.~(\ref{eq:rhovac_t}) and (\ref{eq:tmunu_vac1})), to zeroth order in $h$,
\beq
\rho_{vac} (t) = {\delta G(t) \over G_0} \, \bar{\rho} (t)  \;\;\;\;\;\;\;\;\;\;\;\; 
p_{vac} (t) = w_{vac} \, {\delta G(t) \over G_0} \, \bar{\rho} (t) \; .
\label{eq:rhovac_t_1}
\eeq
with $w_{vac} = \third$.
Therefore, in light of the results of Ref. \cite{hw05}, the problem has been dramatically 
reduced to just computing the much more tractable expression
\beq
- {1 \over 2 \, \nu} \, 
{ 1 \over \Box^{(0)} } \cdot \Box^{(1)} (h) \cdot T_{\mu \nu}^{vac} \; ,
\label{gbox_h_tensor_vac}
\eeq
and in fact the only ordering for which the expression
$ ( \delta G(\Box) / G_0 ) \, T_{\mu \nu} $ is calculable within reasonable effort.
Still, in general the resulting expression for $ { 1 \over \Box^{(0)} } \cdot \Box^{(1)} (h) $
is rather complicated if evaluated for arbitrary functions, although it does
have a structure similar to the one found for the scalar box in Eq.~(\ref{eq:box_ratio}).

Here we will resort, for lack of better insights, to a treatment that parallels what was 
done before for the scalar box, where one assumed a harmonic time dependence for the
metric trace fluctuation $h(t) = h_0 \, e^{i \omega t}$, and similarly for 
$a(t)=a_0 \, e^{i \Gamma t}$ and $\rho (t)= \rho_0 \, e^{i \Gamma t}$.
In the limit $\omega \gg \Gamma$, corresponding to $ \dot{h}  / h  \gg  \dot{a} / a  $, 
one finds for the fluctuation $ \delta \rho_{vac} (t)  $ in Eq.~(\ref{eq:delta_rhovac_t_h})
\beq
\delta \rho_{vac} (t)  =  {\delta G(t) \over G_0}\, \delta \rho (t) + 
{1 \over 2 \, \nu} \, c_h \, {\delta G(t) \over G_0}  \; h(t) \, \bar{\rho}(t) \; .
\eeq
The $O(h)$ correction factor $c_h$ for the tensor box is now given by
\beq
c_h =  {11\over 3} \, \, {\dot{a} \over a}  \, { h \over \dot{h} } \; ,
\label{eq:ch_tens}
\eeq
with all other off-diagonal matrix elements vanishing.
Furthermore one finds to this order, but only for the specific choice 
$w_{vac}=\third$ in the zeroth order $T_{\mu\nu}^{vac}$,  
\beq
\delta p_{vac} (t)  \; = \; \third \, \delta \rho_{vac} (t)  
\eeq
{\it i.e.} the $O(h)$ correction preserves the original result $ w_{vac}= \third $.
In other words, the first order result $O(h)$ just obtained for the tensor box
would have been somewhat inconsistent with the zeroth order result, unless one
had $w_{vac}= \third $ to start with.
Now, one would not necessarily expect that the first order correction could be
still be cast in the form of the same equation of state 
$ p_{vac} \simeq \third \rho_{vac} $ as the the zeroth order result, but 
it would nevertheless seem attractive that such a simple relationship can be
preserved beyond the lowest order.

As far as the magnitude of the correction $c_h$ in Eq.~(\ref{eq:ch_tens}) one
can argue again, as was done in the scalar box case,
that from Eq.~(\ref{eq:ch_fa}) one can relate the combination 
$ ( \dot{h}  / h ) ( a / \dot{a} )$ to the growth index $f(a)$.
Then, in the absence of a running $G$ (which is all that is needed here, to the order
one is working), an explicit form for $f(a)$ is known in terms of suitable
derivatives of a Gauss hypergeometric function. 
These can then be inserted into Eq.~(\ref{eq:ch_tens}).
Alternatively, one can make use again of the fact that for a scale factor referring
to ``today'' $a/a_0 \approx 1$, and for a matter fraction $\Omega \approx 0.25$,
one knows that $ f (a=a_0) \simeq 0.4625 $, and thus in Eq.~(\ref{eq:delta_rhovac_t_h})
$c_h \simeq (11/3) \times 2.1621 = + 7.927 $.
This last result can then be compared to the earlier scalar result which gave 
$c_h \simeq + 2.162 $ using the same set of approximations (slowly varying background fields).
It is encouraging that the new correction is a bit larger but not too different
from what was found before in the scalar box case.
Note that so far the sign of the $O(h)$ correction is the same in all 
physically relevant cases examined.

Next, as in the scalar box case,  one can do the same analysis in the opposite, but
less physical, limit $\omega \ll \Gamma$  or $ \dot{h}  / h  \ll \dot{a} / a  $.
One now obtains from the $tt$ matrix element the $O(h)$  
correction in the expression for $\delta \rho_{vac}$ given in
Eq.~(\ref{eq:delta_rhovac_t_h}), namely
\beq
{1 \over 2 \, \nu} \, c_h \, {\delta G(t) \over G_0}  \; h(t) \, \bar{\rho}(t) \; .
\label{eq:ch_tens_smo}
\eeq
with a coefficient
\beq
c_h =  - {121 \over 60} \, \, { \omega^2 \over \Gamma^2 }  \, = \, 
\simeq - { 121 \over 60 } \, \left ( { a \over \dot{a} }  \right )^2 \, { \ddot h \over h} 
= O ( \ddot{h} / h ) \; .
\label{eq:ch_tens_1t}
\eeq
Similarly for the $ii$ matrix element of the $O(h)$ correction one finds
\beq
{1 \over 2 \, \nu} \, a^2 (t) \, c_h'  \, {\delta G(t) \over G_0}  \; h(t) \, \bar{\rho} (t) \; .
\eeq
with 
\beq
c_h' =  - {5 \over 18} 
\label{eq:ch_tens_1i}
\eeq
giving now the $\delta p_{vac} (h) $ correction.
Again all off-diagonal matrix elements are equal to zero.
It seems therefore that this limit, $\omega \ll \Gamma$  or $ \dot{h}  / h  \ll \dot{a} / a  $,
leads to rather different results compared to
what had been obtained before: the only surviving contribution
to $O(h)$ is a rather large pressure contribution, with a sign that
is opposite to all other cases encountered previously.
Furthermore here the relationship $w_{vac}=\third$ is
no longer preserved to $O(h)$.
But, as emphasized in the previous discussion of the scalar box case,
this second limit is in our opinion less physical, because of the fact 
that now the background is assumed to be varying more rapidly in time 
than the metric perturbation itself,  $ \dot{a} / a  \gg \dot{h}  / h $.
Furthermore, as in the scalar box calculation, one disturbing but not 
entirely surprising general aspect of the whole calculation in this second
$\omega \ll \Gamma$ limit, is its extreme sensitivity as far as magnitudes
and signs of the results are concerned, to the set of 
assumptions initially made about the time development of the background.
As a final sample calculation let us mention here the case, similar to what was
done originally for the scalar box, where one assumes instead 
$ \dot{a} / a \equiv H(a) \approx {\rm const.} $ and $\bar{\rho} \approx {\rm const.} $,
which, as we mentioned previously, seems now a bit restrictive.
Nevertheless we find it instructive to show how sensitive the
calculations are to the nature of the background, and in
particular its assumed time dependence.
In the notation of Eqs.~(\ref{eq:ch_tens_smo}), (\ref{eq:ch_tens_1t})
and (\ref{eq:ch_tens_1i}) one finds in this case
\beq
c_h = + { 625 \over 192 } { \omega^2 \over H^2 } 
=  - { 625 \over 192 } { 1 \over H^2 }  { \ddot{h} \over h} 
\;\;\;\;\;\;\;\; 
c_h' = - { 4 \over 9 }  \; .
\eeq
Again here the pressure contribution $\delta p_{vac} (h) $ is the dominant
contribution, the $\delta \rho_{vac} (h) $ part being negligible, $O (\ddot{h} ) $.
For the reasons mentioned, in the following we will no longer consider 
this limit of rapid background fluctuations any further.

To summarize, the results for a scalar box and for a very slowly varying background,
$ \dot{h}  / h  \gg  \dot{a} / a  $, give the $O(h)$ corrected expression
for $\delta \rho_{vac} (t)$ in Eq.~(\ref{eq:delta_rhovac_t_h}) and 
$\delta p_{vac} (t) = w_{vac} \, \delta \rho_{vac} (t) $ with $c_h \simeq  + 2.162 $,
while the tensor box calculation, under essentially the same assumptions,
gives the somewhat larger result $c_h \simeq  + 7.927 $.
From now on, these will be the only two choices we shall consider here.

\vskip 40pt
\subsection{First Order Energy-Momentum Conservation}
\hspace*{\parindent}
\label{sec:enmom_first}

The next step in the analysis involves the derivation of the energy-momentum conservation
to first order in the fluctuations, and a derivation of the relevant field equations to the same order.
After that, energy conservation will be used to eliminate the $h$ field entirely, and thus
obtain a single equation for the matter density fluctuation $\delta$.

The results so far can be summarized as follows.
For the metric in Eq.~(\ref{eq:pert_metric}), and in the limit ${\bf q} \rightarrow 0$, 
the field equations in Eq.~(\ref{eq:field1}) can now be written as
as
\beq
R_{\mu\nu} \, - \, \half \, g_{\mu\nu} \, R \, + \, \lambda \, g_{\mu\nu}
\; = \; 8 \pi \, G_0 \left ( T_{\mu\nu}   +  T_{\mu\nu}^{vac} \right ) \; ,
\label{eq:field3}
\eeq
with $ T_{\mu\nu}^{vac} \, \equiv \, ( \delta G(\Box) / G_0 ) \, T_{\mu\nu} $.
Here  $T_{\mu\nu}$ describes the ordinary matter contribution, in the form of a perfect fluid 
as given in Eq.~(\ref{eq:tmunu_perf}), here with $p=w \rho$ and $w \simeq 0$, 
while $T_{\mu\nu}^{vac}$ describes the additional vacuum polarization contribution
\beq
T_{\mu \nu}^{vac} = \left [ \, p_{vac} (t) + \rho_{vac} (t) \, \right ] u_\mu \, u_\nu 
+ g_{\mu \nu} \, p_{vac} (t)
\eeq
with $p_{vac}=w_{vac} \, \rho_{vac}$ and $w_{vac}=\third$, as in Eq.~(\ref{eq:wvac_def}).
Furthermore, each field now contains both a background and a perturbation contribution,
\beq
\rho(t) = \bar{\rho} (t) + \delta \rho (t)     
\;\;\;\;\;\;\;\;\;\;\;\;
p (t) = w \, \rho (t) \; 
\eeq
and similarly
\beq
\rho_{vac} (t) = \bar{\rho}_{vac} (t) + \delta \rho_{vac} (t)     
\;\;\;\;\;\;\;\;\;\;\;\;
p_{vac} (t) = w_{vac} \, \rho_{vac} (t) \; .
\eeq
From Eq.~(\ref{eq:rhovac_t}) one has
\beq
\bar{\rho}_{vac} (t) = {\delta G(t) \over G_0} \, \rho (t)  \; ,
\eeq
while  from Eq.~(\ref{eq:delta_rhovac_t_h}) on has
\beq
\delta \rho_{vac} (t)  =  {\delta G(t) \over G_0}\, \delta \rho (t) + 
{1 \over 2 \, \nu} \, c_h \, {\delta G(t) \over G_0}  \; h(t) \, \bar{\rho}(t)
\label{eq:delta_rhovac_t_h_1}
\eeq
and similarly $ \delta p_{vac} (t) =  w_{vac}  \, \delta \rho_{vac} (t)  $.
The second $O(h)$ terms in both expressions physically account for the
feedback of the metric fluctuations $h$ on the vaccum density 
$\delta \rho_{vac}$ and pressure $\delta p_{vac}$ fluctuations.
In light of the discussion of the previous section, we will limit our derivations
below to the case of constant  $c_h$; the case of a non-constant $c_h$ 
as in Eq.~(\ref{eq:ch_fa}) can be dealt with as well, but the resulting equations
are found to be quite a bit more complicated to write down.

Consequently all quantities in the effective field equations of Eq.~(\ref{eq:field3})
have been specified to the required order in the field perturbation expansion.
First we will look here at the implications of energy-momentum conservation,
$ \nabla^\mu \left ( T_{\mu\nu}   +  T_{\mu\nu}^{vac}  \right ) = 0 $, 
to first order in the fluctuations.
The zeroth order energy conservation equation was already obtained in
Eq.~(\ref{eq:encons_zeroth_w}), and its explicit solution 
for $\bar{\rho} (a) $ given in Eq.~(\ref{eq:rho_zeroth}).
After defining the matter density contrast $\delta (t)$  as the ratio
$ \delta (t)  \equiv \delta \rho (t) /  \bar{\rho} (t) $,
the energy conservation equation to first order in the perturbations
is found to be 
\bea
\left [ 
\, - {1 \over 2} \left ( \left ( 1+ w \right ) + 
\left ( 1 + w_{vac} \right ) \, {\delta G(t) \over G_0} \right ) - 
{1 \over 2 \nu} \, c_h \, {\delta G(t) \over G_0} \,
\right ] \, & \dot{h} (t) &
\nonumber \\
 + \,\,\,\, 
\left [ 
\, {1 \over 2 \nu} \, c_h \, 
\left ( 3 \left ( w - w_{vac} \right ) \, {\dot{a} (t) \over a (t)} \, {\delta G(t) \over G_0}
- { \dot{\delta G}(t) \over G_0} \right ) 
\right ] & h(t) &
 = \,\,\, \left [ 1 + {\delta G(t) \over G_0} \right ] \,  \dot{\delta} (t) \; .
\label{eq:encons_fluc}
\eea
In the absence of a running $G$ ($\delta G(t) =0$) this reduces simply to
$ - \half \, \left ( 1 + w \right )\, \dot{h} (t) = \dot{\delta} (t) $,
and thus to the standard result for the metric trace perturbation
in terms of the density contrast
\beq
- \half \, \left ( 1 + w \right )\, h (t) = \delta (t) \; .
\label{eq:hdelta}
\eeq
This last result then allows us to solve explicitly, at the given order, 
{\it i.e.} to first order
in the fluctuations and to first order in $\delta G$, for the metric perturbation
$ \dot{h} (t) $ in terms of the matter density fluctuation 
$ \delta (t)  $ and $ \dot{\delta} (t) $,
\bea
\dot{h} (t) = & - & \, 
{ 2 \over  1 + w } \left [ 
\, 1 + { 1 \over  1 + w  } \,  \left ( \left (w -w_{vac} \right ) 
- 2 \, c_h \, {1 \over 2 \, \nu } \right ) \, 
{ \delta G (t) \over G_0 } \,
\right ] \, \dot{\delta} (t) 
\nonumber \\
& - & \, {1 \over 2 \, \nu} \, { 4 \, c_h \over \left (1 + w \right )^2 } 
\left [ 
\, 3 \,\left (w - w_{vac} \right ) \, { \dot{a} (t) \over a (t) } \,  
{ \delta G (t) \over G_0 } \, 
-  \, { \dot{\delta G} (t) \over G_0 } \,
\right ] \delta (t) \; .
\label{eq:hdotdelta}
\eea
Similarly, by differentiating the above relationship, an expression for 
$ \ddot{h} (t) $ in terms of $\delta$ and its derivatives can be
obtained as well.

\vskip 40pt
\subsection{First Order Field Equations}
\hspace*{\parindent}
\label{sec:field_first}

To first order in the perturbations, the $tt$ and $ii$ effective field equations 
become, respectively,
\beq
{\dot{a} (t) \over a (t)}\, \dot{h} (t)  -  
8 \pi \, G_0 \, {1 \over 2 \nu} \, c_h \,  {\delta G(t) \over G_0}\, \bar{\rho}(t) \, h(t)
=  8 \pi \, G_0 \left ( 1 +  {\delta G(t) \over G_0} \right ) \, 
\bar{\rho} (t) \, \delta (t) 
\label{eq:field_fluc_tt}
\eeq
and
\beq
\ddot{h} (t)\, + \, 3 \,  {\dot{a} (t) \over a (t)}\, \dot{h} (t) \,  
+ \, 24 \pi \, G_0 \, {1 \over 2 \nu} \, c_h \, w_{vac} \, 
{\delta G(t) \over G_0}\, \bar{\rho} (t) \, h (t) 
=   - \, 24 \pi \, G_0 \left ( w + w_{vac}\,{\delta G(t) \over G_0} 
\right ) \, \bar{\rho}(t)\, \delta (t)
\label{eq:field_fluc_ii}
\eeq
In the second $ii$ equation, the zeroth order $ii$ field equation
of Eq.~(\ref{eq:fried_run}) has been used to achieve some simplification.

As a final exercise, it is easy to check the overall consistency of the first order 
energy conservation 
equation of Eq.~(\ref{eq:encons_fluc}),  and of the two field equations given in
Eqs.~(\ref{eq:field_fluc_tt}) and (\ref{eq:field_fluc_ii}).
To do so, one needs to 
(i) take the time derivative of the $ tt $ equation in Eq.~(\ref{eq:field_fluc_tt});
(ii) get rid of  $ \dot{\bar{\rho}} $  consistently by using energy conservation to
zeroth order in $ \delta G $ and in the fluctuations 
 from Eq.~(\ref{eq:encons_fluc}) for terms of order $\delta G$ times a fluctuation,
combined with the use of
energy conservation to first order in $ \delta G $  but without fluctuations 
as in Eq.~(\ref{eq:encons_zeroth}) for the terms that are already of
first order in the fluctuations;
(iii) eliminate the $ \dot{\delta} $ terms using the energy conservation 
equation to first order in  $  \delta G $ without field fluctuations 
(Eq.~(\ref{eq:encons_zeroth})) for  terms proportional to $ \delta G $
times a fluctuation,  and using the energy conservation equation 
to first order in  $  \delta G $ and in the fluctuation 
(again Eq.~(\ref{eq:encons_fluc})) for terms of zeroth order in the fluctuations;
(iv) use the combination of Eqs.~(\ref{eq:fried_run}) that does not
contain $\lambda$, Eq.~(\ref{eq:fried_run_nolambda}), to get rid of
$ \ddot{a} / a $ terms;
(v) Finally use the $tt$ equation for the fluctuation, Eq.~(\ref{eq:field_fluc_tt}),
to eliminate some terms proportional to $ {\bar{\rho}} $ times a fluctuation
so as to finally obtain the second $ii$ field  equation Eq.~(\ref{eq:field_fluc_ii}).

\vskip 40pt
\subsection{Matter Density Contrast Equation in ${\bf t}$}
\hspace*{\parindent}
\label{sec:delta_eq}

To obtain an equation for the matter density contrast 
$\delta (t) = \delta \rho (t) / \bar{\rho} (t)$
one needs to eliminate the metric trace field $h(t)$ from the field equations.
This is first done by taking a suitable linear combination of the two field equations 
in Eqs.~(\ref{eq:field_fluc_tt}) and (\ref{eq:field_fluc_ii}), to get the
equivalent equation
\bea
\ddot{h} (t) + 2 \, {\dot{a} (t) \over a (t)} \, \dot{h} (t) 
& + & 8 \pi \, G_0 \, {1 \over 2 \nu} \, c_h \,
\left ( 1 + 3 \, w_{vac} \right ) 
{\delta G(t) \over G_0} \, \bar{\rho}(t) \, h(t) 
\nonumber \\
& = & - \, 8 \pi \, G_0 \left [
 \left ( 1 + 3 \, w \right ) + 
\left ( 1 + 3 \, w_{vac} \right ) \,  {\delta G(t) \over G_0}
\right ] \bar{\rho} (t) \, \delta (t) \; .
\label{eq:field_fluc_h}
\eea
Then the first order energy conservation equations to zeroth 
(Eq.~(\ref{eq:hdelta})) and first (Eq.~(\ref{eq:hdotdelta})) order
in $\delta G$ allow one to completely eliminate the $h$, $ \dot{h} $ and
$ \ddot{h} $ field in terms of the matter density perturbation
$\delta (t)$ and its derivatives.
The resulting equation reads, for $ w = 0 $ and $w_{vac}= \third $,
 \bea
\ddot{\delta}(t) && 
+ \left [ 
\left ( 2 \, {\dot{a}(t) \over a(t)}
- {1 \over 3} \, {\dot{\delta G}(t) \over G_0} \right ) 
- {1 \over 2 \nu} \cdot 2 \, c_h \cdot \left ( {\dot{a}(t) \over a(t)} \, {\delta G(t) \over G_0} 
+ 2\,{\dot{\delta G}(t) \over G_0} \right ) 
\right ] \dot{\delta}(t) 
\nonumber \\
&& + \left [
- \, 4 \pi \, G_0  \left (1 + {7 \over 3} \, {\delta G(t) \over G_0} 
- {1 \over 2 \nu} \cdot 2 \, c_h \cdot {\delta G(t) \over G_0} \right ) \bar{\rho}(t) 
\right. 
\nonumber \\
&& \;\;\;\;\;  
\left.  
- {1 \over 2 \nu} \cdot 2 \, c_h \cdot 
\left ( {{\dot{a}}^2(t) \over a^2(t)} \, {\delta G(t) \over G_0} 
+ 3 \, {\dot{a}(t) \over a(t)} \, {\dot{\delta G}(t) \over G_0} 
+ {\ddot{a}(t) \over a(t)} \, {\delta G(t) \over G_0} 
+ {\ddot{\delta G}(t) \over G_0} \right ) 
\right ] \delta(t) = 0 \; .
\nonumber \\
\label{eq:dfq_delta_Gbox}
\eea
This last equation then describes matter density perturbations to linear order,
taking into account the running of $ G(\Box) $, and is therefore the main result
of this paper.
The terms proportional to $c_h$, which can be clearly identified in the above
equation, describe the feedback of the metric fluctuations $h$ on the 
vaccum density $\delta \rho_{vac}$ and pressure $\delta p_{vac}$ fluctuations.
The equation given above can now be compared with the corresponding, much simpler, 
equation obtained for constant $G$,  {\it i.e.,} for $ G \rightarrow G_0 $ and still $ w = 0 $ 
(see for example \cite{wei72} and \cite{pee93})
\beq
\ddot{\delta} (t) + 2 \, {\dot{a} \over a} \, \dot{\delta} (t) 
- 4 \pi \, G_0 \, \bar{\rho}(t) \, \delta(t) = 0 \; 
\label{eq:dfq_delta_G0}
\eeq
from which one obtains for the growing mode
\beq
\delta_{\bf q} (t) = \delta_{\bf q} ( t_0) \, \left ( { t \over t_0 } \right )^{2/3} \; ,
\eeq
which is the standard result in the matter-dominated era.

\vskip 40pt
\subsection{Matter Density Contrast Equation in ${\bf a(t)}$}
\hspace*{\parindent}
\label{sec:delta_eq_a}

It is common practice at this point to write an equation for the density contrast 
$\delta(a)$ as a function not of $t$, but of the scale factor $a(t)$.
This is done by utilizing the following simple derivative identities
\beq
\dot{f}(t) = a \, H (a) \, {\partial f (a) \over \partial a}
\label{eq:f_dot}
\eeq
\beq
\ddot{f} (t) = {a}^2 \,  H^2 (a) 
\left ( {\partial \ln H (a) \over \partial a} + {1 \over a} \right ) 
\, {\partial f (a) \over \partial a} 
+ {a}^2 \, H^2 (a) \, {{\partial}^2 f (a) \over \partial {a}^2}
\label{eq:f_ddot}
\eeq
where $ f $ is any function of $ t $, and $ H \equiv \dot{a} (t) / a (t) $ the Hubble constant.
This last quantity can be obtained from the zeroth order $tt$ field equation
\beq
H^2 (a) \equiv \left ( { \dot{a} \over a } \right )^2 = { 8 \, \pi \, G_0 \over 3 } \, \bar{\rho} 
+ { \lambda \over 3 } \; .
\eeq
Often this last equation is written in terms of current density fractions,
\beq
H^2 (a) \equiv \left ( {\dot{a} \over a} \right )^2 
= \left ( \dot{z} \over 1 + z \right )^2 
= H_0^2 \left [ \Omega \, \left ( 1 + z \right )^3 
+ \Omega_R \, \left ( 1 + z \right )^2 + \Omega_{\lambda} \right ]
\eeq
with $ a/a_0 = 1 / ( 1 + z ) $ where $ z $ is the red shift, and $a_0$ the scale
factor ``today''.
Then $ H_0 $ is the Hubble constant evaluated today, 
$ \Omega $ the (baryonic and dark) matter density, $ \Omega_R $ the space curvature
contribution corresponding to a curvature $ k $ term, and $\Omega_{\lambda} $
the dark energy or cosmological constant part, all again measured {\it today}.
In the absence of spatial curvature $k=0$ one has today
\beq
\Omega_{\lambda} \equiv {\lambda \over 3 \, H_0^2}
\;\;\;\;\;\;
\Omega \equiv { 8 \, \pi \, G_0 \, \bar{\rho}_0 \over 3 \, H_0^2 } 
\;\;\;\;\;\;\;\;\;\; 
\Omega + \Omega_{\lambda} = 1 \; .
\label{eq:omega_def}
\eeq
In terms of the scale factor $a(t)$ the equation for matter density
perturbations for constant $G=G_0$, Eq.~(\ref{eq:dfq_delta_G0}), becomes
\beq
{\partial^2 \delta(a) \over \partial a^2 } 
+ \left [ {\partial \ln H(a) \over \partial a} 
+ {3 \over a } \right ] \, {\partial \, \delta(a) \over \partial a} 
- 4 \pi \, G_0 \,{1 \over a^2 H(a)^2}\, \bar{\rho} (a) \, \delta (a) = 0 \; .
\eeq
The quantity $H(a)$ is most simply obtained from the FLRW field equations
\beq
H (a) = \sqrt{ {8 \pi \over 3} \,  G_0 \, \bar{\rho} (a) + {\lambda \over 3} } \; ,
\label{eq:Ha}
\eeq
with the matter density given in Eq.~(\ref{eq:rho_fried}), 
which can in principle be solved for $a(t)$,
\beq
t- t_0 = \int
{ da \over a \, \sqrt{ 
{8 \pi \over 3} \,  G_0 \, \bar{\rho}_0 \, \left ({a_{0} \over a} \right )^3  
+ {\lambda \over 3}  } } \; .
\label{eq:a_t}
\eeq
It is convenient at this stage to introduce a parameter $\theta$ describing the cosmological 
constant fraction as measured today,
\beq
\theta \equiv { \lambda \over 8 \, \pi \, G_0 \, \bar{\rho}_0 }
\, =  \, { \Omega_{\lambda} \over \Omega}  \, = \,  { 1 - \Omega \over \Omega }  \; .
\label{eq:theta_def}
\eeq
While the following discussion will continue with some level of generality, in practice
one is mostly interested in the observationally favored case of
a current matter fraction $\Omega \approx 0.25$, for which $\theta \approx 3$.
In terms of the parameter $\theta$ the equation for the density contrast $\delta (a)$ for constant 
$G$ can then be recast in the slightly simpler form 
\beq
{\partial^2 \delta(a) \over \partial a^2 }
+  { 3  \, ( 1 + 2 \, a^3 \, \theta ) \over 2 \, a \, ( 1 + a^3 \, \theta  ) }  \,
{\partial \, \delta(a) \over \partial a} 
- { 3 \over 2 \, a^2 \, ( 1 + a^3 \, \theta ) } \, \delta (a) = 0 \; .
\label{eq:dfq_delta_G0_a}
\eeq
A general solution of the above equation is given
by a linear combination of the two solutions
\beq
\delta_0 (a) =
c_1 \cdot  \sqrt{1 + a^3 \, \theta } \,  a^{- 3 / 2}  + 
c_2 \cdot a \cdot {}_2 F_1 \, \left ( {1 \over 3}, 1; {11 \over 6}; - a^3 \, \theta \right )  
\eeq
where  $ c_1 $ and $ c_2 $ are arbitrary constants, and ${}_2 F_1$ is the  Gauss
hypergeometric function.
The subscript $0$ in $\delta_0 (a)$ is to remind us that 
this solution  is appropriate for the case of constant $G=G_0$.
Since one is only interested in the growing solution, the constant $c_1=0$.

To evaluate the correction to $\delta_0 (a) $ coming from the terms proportional to
$c_a$ one sets
\beq
\delta (a)  \propto \delta_0  (a) \, \left [ \, 1 + c_a \, {\cal F} (a) \, \right ] \; ,
\label{eq:fa_corr}
\eeq
and inserts the resulting expression in Eq.~(\ref{eq:dfq_delta_Gbox}), written now as
a differential equation in $a (t)$, after using Eqs.~(\ref{eq:f_dot}) and (\ref{eq:f_ddot}) 
to replace
\bea
\dot{a} (t)& = & a\, H \nonumber \\
\ddot{a} (t)& = & {a}^2 \, {H}^2 \left ( {\partial \ln H \over \partial a} 
+ {1 \over a} \right ) \; .
\label{eq:a_ttoa}
\eea
One only needs to determine the differential equations for density perturbations 
$ \delta $ up to first order in the fluctuations, so it will be sufficient to 
obtain an expression for Hubble constant $ H $ from the 
$ tt $ component of the effective field equation to zeroth 
order in the fluctuations, namely the first of Eqs.~(\ref{eq:fried_run}).
One has
\beq
H (a) = \sqrt{{8 \pi \over 3} \,  G_0 \left ( 1 
+ {\delta G (a) \over G_0} \right ) \, \bar{\rho} (a) + {\lambda \over 3}}
\label{eq:H_in_efe}
\eeq
with $G(a) $ given in Eq.~(\ref{eq:grun_a})
and $\bar{\rho} (a)$ given in Eq.~(\ref{eq:rho_zeroth_sim}).
\footnote{
We have noted before that Eq.~(\ref{eq:H_in_efe}) is suggestive of a 
small additional matter contribution, 
$\Omega_{vac} \simeq (8 \pi /3 ) \delta G (a) \bar{\rho}_0 / H_0^2  $,
to the overall balance in Eq.~(\ref{eq:omega_def}).}
In this last expression the exponent is $ \gamma_\nu = 3/2 \nu \simeq 9/2$ for
a matter dominated background universe, although more general choices, such
as $ \gamma_\nu = 3 (1+w)/2 \nu $ or even the use of Eq.~(\ref{eq:a_t}), 
are possible and should be explored (see discussion later).
Also, $c_a \approx c_t$ if $a_0$ is identified with a scale factor 
corresponding to a universe of size $\xi$; to a good approximation this corresponds
to the universe ``today'',  with the relative scale factor customarily normalized at
that time to $a/a_0=1$.
In \cite{hw05} it was found that in Eq.~(\ref{eq:grun_t}) $c_t \simeq 0.785 \, c_0 $ 
in the scalar box case, and $c_t \simeq 0.450 \, c_0 $ in the tensor box case;
in the following we will use the average of the two values.

After the various substitutions and insertions have been performed, one obtains,
after expanding to linear order in $a_0$, a second order linear differential equation
for the correction ${\cal F} (a)$  to $\delta (a)$, as defined in Eq.~(\ref{eq:fa_corr}).
Since this equation looks rather complicated for general $\delta G(a)$ it won't be 
recorded here, but it is easily obtained from Eq.~(\ref{eq:dfq_delta_Gbox}) 
by a sequence of straightforward substitutions and expansions.
The resulting equation can then be solved for ${\cal F} (a)$,
giving the desired density contrast $\delta (a)$ as a function of the 
parameter $\Omega$.

Nevertheless with the specific choice for $G(a)$ given in Eq.~(\ref{eq:grun_a})
an explicit form for the equation for $\delta (a)$ reads:
\beq 
{\partial^2 \delta (a) \over \partial a^2 } 
+  A (a) \, {\partial \delta (a) \over \partial a} + B(a) \, \delta (a) = 0 \; .
\label{eq:dfq_delta_Gbox_a}
\eeq
with the two coefficients given by
\bea
&& A(a)  = 
\frac{3 ( 1 + 2 a^3 \theta ) }{ 2 a ( 1 + a^3 \theta ) } -
\nonumber \\
 && \frac{c_a \left( 9 a^3 \! \left( 1 \! + \! \gamma_\nu  \right)  
        \theta  \nu  \! + \! a^{\gamma_\nu } \!
        \left( 6 c_h \gamma_\nu  \left( 1 \! + \! 2 \gamma_\nu  \right)  \!
           {\left( 1 \! + \! a^3 \theta  \right) }^2 \! + \! 
          \left( -9 a^3 \theta  \! + \! 
             \gamma_\nu  \left( 1 \! + \! a^3 \theta  \right)  \!
              \left( 3 \! + \! 2 \gamma_\nu  \left( 1 \! + \! a^3 \theta  \right) 
                \right)  \right) \! \nu  \right)  \right) } 
      { 6 \, a \, \nu \, \gamma_\nu  \,
        {\left( 1 \! + \! a^3 \theta  \right) }^2 }
\nonumber \\
\eea
and
\bea
&& B(a) = - \frac{3}{2 a^2 (1+  a^3 \theta ) }  -
\nonumber \\
&& \frac{c_a \left( 3 a^3 \left( 1 \! + \! \gamma_\nu  \right)  
        \theta  \nu  \! + \! a^{\gamma_\nu }  \!
        \left( c_h \gamma_\nu  \left( 2 \! + \! \gamma_\nu  \right)  \!
           \left( 1 \! + \! a^3 \theta  \right)  \!
           \left( -1 \! + \! 2 \gamma_\nu  \! + \! 
             2 a^3 \! \left( 1 \! + \! \gamma_\nu  \right)  \theta  \right)  \! + \! 
          \left( 4 \gamma_\nu  \! + \! 
             a^3 \! \left( -3 \! + \! 4 \gamma_\nu  \right)  \theta  \right)  \! \nu 
          \right)  \right) }
     {2 \, \nu \, \gamma_\nu  \, a^2 \, 
       {\left( 1 + a^3 \theta  \right) }^2 }
\nonumber \\
\nonumber \\
\eea
and the variable $a$ considered just as a stand-in for what should really be the variable $a/a_0$.
To obtain an explicit solution to the above equation one needs to know the coefficient
$c_a$ and the exponent $\gamma_\nu$ in Eq.~(\ref{eq:grun_a}), whose likely values are
discussed above and right after the quoted expression for $G(a)$. 
For the exponent $\nu$ one has $\nu \simeq \third$, whereas for the value for $c_h$ one
finds, according to the discussion in the previous sections, 
$c_h \simeq 7.927 $ for the tensor box case.
Furthermore one needs at some point to insert a value for the matter density fraction
parameter $\theta$ as given in Eq.~(\ref{eq:theta_def}), which based on current
observation is close to $\theta = ( 1 - \Omega) / \Omega \simeq 3 $.


\vskip 40pt
\newsection{Relativistic Growth Index with ${\bf G(\Box)}$}
\hspace*{\parindent}
\label{sec:growth}

The solution of the above differential equation for the matter density contrast in the 
presence of a running Newton's constant $G(\Box)$ leads to an explicit
form for the function $\delta (a) = \delta_0 (a) [ 1 + c_a {\cal F} (a) ] $.
From it, an estimate of the size of the corrections coming from the new terms
due to the running of $G$ can be obtained.
It is clear from the previous discussion, and the form of $G(\Box)$, that
such corrections will become increasingly important in the present
era $t \approx t_0 $ or $a \approx a_0 $.
When discussing the growth of density perturbations in classical
General Relativity it is customary at this point to introduce a 
scale-factor-dependent {\it growth index} $ f (a) $ defined as 
\beq
f (a) \equiv { \partial \ln \delta (a) \over \partial \ln a }  \; ,
\label{eq:fa_def}
\eeq
which is in principle obtained from the differential equation for any scale factor $a(t)$.
Nevertheless, here one is mainly interested in the neighborhood
of the present era, $a (t) \approx a_0$.
One therefore introduces today's {\it growth index parameter} $ \gamma $ via
\beq
f (a=a_0) \equiv \left. { \partial \ln \delta (a) \over \partial \ln a } 
\right \vert_{a=a_0} \equiv \; \Omega^{\gamma} \; ,
\eeq
so that the exponent $\gamma$ itself is obtained via
\beq
\gamma \equiv  \left. { \ln f \over \ln \Omega  }  \right \vert_{a=a_0} \; .
\label{eq:gamma_def}
\eeq
The solution of the above differential equation for $\delta (a)$ then determines an explicit 
value for the growth index $\gamma$ parameter, for any value of the current
matter fraction $\Omega$.
In the end, because of observational constraints, one is mostly interested in the range
$\Omega \approx 0.25$, so the following discussion will be limited to this case only,
although from the original differential equation for $\delta (a)$
one can in principle obtain a solution for any sensible $\Omega$.
Numerically the differential equation for $\delta (a)$ can in principle be solved
for any value of the parameters.
In practice we have found it convenient, and adequate, to obtain the solution
as a power series in either $\Omega$ or $1-\Omega$. 
In the first case the resulting series is asymptotic and only slowly convergent
around $\Omega \approx 0.25$, while in the latter case the convergence is
much more rapid.
In this last case we have carried therefore the expansion up to eighth order, which gives
the answers given below (see also Figures 1-4) to an accuracy of several decimals.

It is known that in the absence of a running Newton's constant $G$ 
($ G \rightarrow G_0 $, thus $c_a=0$) one has 
$ f (a=a_0) = 0.4625 $ and $ \gamma = 0.5562 $ for the standard
$\Lambda CDM$ scenario with $ \Omega = 0.25 $ \cite{pee93}.
On the other hand, when the running of $G (\Box)$ is taken into account, 
one finds from the solution to Eq.~(\ref{eq:dfq_delta_Gbox}) for the growth 
index parameter $ \gamma $ at $\Omega =0.25 $
the following set of results.

For the tensor box case discussed in Sec.~(\ref{sec:t_box}) one has
the value $c_h= (11/3) \times 2.1621 = 7.927 $ in 
Eqs.~(\ref{eq:delta_rhovac_t_h}) and (\ref{eq:delta_rhovac_t_h_1}), 
which gives
\beq 
\gamma =  0.5562 - 199.2 \, c_a + O(c_a^2 ) \; .
\label{eq:gamma_relat_ten}
\eeq
For the scalar box case discussed in Sec.~(\ref{sec:s_box}) one has instead
$c_h= 2.1621 $ and in this case one finds
\beq 
\gamma =  0.5562 - 54.8 \, c_a + O(c_a^2 ) \; .
\label{eq:gamma_relat_sca}
\eeq
As a comparison, we have also computed the exponent $\gamma$ for the case $c_h=0$
in Eqs.~(\ref{eq:delta_rhovac_t_h}) and (\ref{eq:delta_rhovac_t_h_1}).
This corresponds to a case where the $O(h)$ correction to $\delta \rho_{vac}$
is entirely neglected, and one obtains $ \gamma =  0.5562 - 0.703 \, c_a + O(c_a^2 ) $.
Finally for the Newtonian (non-relativistic) treatment, described in Appendix A, one finds
the much smaller correction
\beq 
\gamma =  0.5562 - 0.0142 \, c_a + O(c_a^2 ) \; .
\label{eq:gamma_newt}
\eeq
Among these last expressions, the tensor box case is supposed to give ultimately the correct answer;
the scalar box case only serves as a qualitative comparison, and the $c_h=0$ case is done
to estimate independently the size of the correction coming from the ubiquitous $O(h)$ 
or ${1 \over 2 \nu} \, c_h$ terms (see for example the differential equation for the
density perturbations $\delta (t)$ in Eq.~(\ref{eq:dfq_delta_Gbox})).
Note that the $c_h=0$, scalar and tensor box results can be summarized
into the slightly more general formula
\beq 
\gamma =  0.5562 - ( 0.703 + 25.04 \, c_h ) \, c_a + O(c_a^2 ) \; .
\label{eq:gamma_ch}
\eeq
showing again the overall importance of the $c_h$ contribution to $\delta \rho_{vac}$
in Eq.~(\ref{eq:delta_rhovac_t_h}).
This last term is responsible for the feedback of the metric fluctuations $h$ on the 
vacuum density $\delta \rho_{vac}$ and pressure $\delta p_{vac}$ fluctuations.

It should be emphasized here once again that all of the above results have been
obtained by solving the differential equation for $\delta (a) $,
Eq.~(\ref{eq:dfq_delta_Gbox_a}), with $G(a)$ given in Eq.~(\ref{eq:grun_a}),
and exponent $ \gamma_\nu = 3/2 \nu \simeq 9/2$ relevant for
a matter dominated background universe.
It is this last choice that needs to be critically analyzed, as it might give
rise to a definite bias.
Our value for $ \gamma_\nu $ so far reflects our 
choice of a matter dominated background.
More general choices, such as an ``effective'' $\gamma_\nu = 3 (1+w)/2 \nu $ 
with and ``effective'' $w$, or even the use of Eq.~(\ref{eq:a_t}), 
are in principle possible.
Then, although Eq.~(\ref{eq:dfq_delta_Gbox}) for $\delta (t)$ remains unchanged, 
Eq.~(\ref{eq:dfq_delta_Gbox_a}) for $\delta (a)$ would have to be solved with new 
parameters.
In the next section we will discuss a number of options which should allow one
to increase on the accuracy of the above result, and in particular correct
the possible shortcomings coming so far from the specific choice of the exponent 
$\gamma_\nu$.

\vskip 40pt
\subsection{Possible Physical Interpretation of the Results}
\hspace*{\parindent}
\label{sec:inter}

Looking at these last results (see also Figs. 1-4), they seem to indicate that (a) the correction due to 
the $h$ (or $1/2 \nu $) terms in Eq.~(\ref{eq:delta_rhovac_t_h}) and in the differential
equation, Eq.~(\ref{eq:dfq_delta_Gbox}), for $\delta (a)$ 
is rather large, and that (b) it is more than twice as large in the tensor box case 
than it is in the scalar box case. 
Furthermore they seem to suggest that (c) the Newtonian
(non-relativistic) result, which does not contain a $\rho_{vac}$ contribution,
substantially underestimates the size of the quantum correction.
To quantitatively estimate the actual size of the correction in the above expressions
for the growth index parameter $\gamma$, and make some preliminary comparison
to astrophysical observations, some additional information is needed.

The first item is the coefficient $c_0 \approx 33.3$ in Eq.~(\ref{eq:grun_box})
as obtained from lattice gravity calculations of invariant correlation functions at 
fixed geodesic distance \cite{cor94}.
We have re-analyzed the results of \cite{cor94} which involve rather large uncertainties for
this particular quantity,
nevertheless it would seem difficult to accommodate values for $c_0$ that are
more than an order of magnitude smaller than the quoted value.
A renewed more accurate lattice calculation of $c_0$, obtained from the computation
of invariant curvature correlation functions at fixed geodesic distance,
would seem rather desirable at this point.

The next item that is needed here is a quantitative estimate for the magnitude of 
the coefficient $c_a$ in Eq.~(\ref{eq:grun_a}) in terms of $c_t$ in Eq.~(\ref{eq:grun_t}),
and therefore in terms of $c_0$ in the original Eq.~(\ref{eq:grun_box}).
First of all one has $c_a \approx c_t$, if $a_0$ is identified with a scale factor 
corresponding to a universe of size $\xi$; to a good approximation this corresponds
to the universe ``today'',  with the relative scale factor customarily normalized at
that time to $a/a_0=1$, although some large conversion factor might be hidden
in this perhaps naive identification (see below).

Regarding the numerical value of the coefficient $c_t$ itself, it was found in \cite{hw05} that in 
Eq.~(\ref{eq:grun_t}) $c_t \simeq 0.785 \, c_0 $ in the scalar box case, 
and $c_t \simeq 0.450 \, c_0 $ in the tensor box case.
In both cases these estimates refer to values obtained from the zeroth 
order covariant effective field equations.
In the following we will take for concreteness the average of the two values, 
thus $c_t \approx 0.618 \, c_0 $.
Then for all three covariant calculations recorded above
$c_a \approx 0.618 \times 33.3 \approx 20.6 $,
a rather large coefficient.

From all of these considerations one would tend to get estimates for 
the growth parameter $\gamma$ with rather large corrections!
For example, in the tensor box case the corrections would add up to
$ - 199. \, c_a = - 199. \times 0.618 \times 33.3 = - 4095. $.
Even in the Newtonian (non-relativistic) case, where the correction is found
to be the smallest, the corresponding result appears to be quite large.
In this last case $c_a \approx c_t \approx 2.7 \, c_0$ (see Appendix A), so 
the correction to the index $\gamma$ becomes 
$ - 0.0142 \times 2.7 \times 33.3 = - 1.28 $.

It would seem though that one should account somewhere for the fact that the largest 
galaxy clusters and superclusters studied today up to redshifts $ z \simeq 1 $ 
extend for only about, at the very most, $1/20$ the overall size of the visible universe.
This would suggest then that the corresponding scale for the running coupling 
$G (t)$ or $G(a)$ in Eqs.~(\ref{eq:grun_t}) and (\ref{eq:grun_a}) respectively, 
should be reduced by a suitable ratio of the two relevant length scales, 
one for the largest observed galaxy clusters or superclusters, and the second
for the very large, cosmological scale 
$\xi \sim 1 / \sqrt{\lambda/3} \sim 1.51 \times 10^{28} {\rm cm} $ entering the expression
for $\delta G (\Box) $ in Eqs.~(\ref{eq:field1}) and (\ref{eq:grun_box}).
This would dramatically reduce the magnitude of the quantum correction by as much as a factor 
of the order of $(1/20)^{\gamma_{\nu}} = (1/20)^{4.5} \approx 1.398 \times 10^{-6}$.
When this correction factor is roughly taken into account, one obtains the more 
reasonable (and perhaps observationally more compatible) estimates for the
tensor box case 
\beq 
\gamma =  0.5562 - 0.0057  \, c_a + O(c_a^2 ) \; .
\label{eq:gamma_relat_ten_1}
\eeq
and for the scalar box case
\beq 
\gamma =  0.5562 - 0.0016  \, c_a + O(c_a^2 ) \; .
\label{eq:gamma_relat_sca_1}
\eeq
while in the non-relativistic (Newtonian) case one finds
$ \gamma \approx  0.5562 - 4.08 \times 10^{-7}$.
In the tensor box case this would then amount to a slightly reduced value for 
the growth index $\gamma$ {\it at these scales} as compared to the constant $G$ case, 
by as much as a few percent, which could perhaps be observable in the 
not too distant future.
Of course, on larger scales the effects would be more significant, and
somewhat bigger for larger values of $\Omega$.

A second possibility we will pursue here briefly is to consider a shortcoming, 
mentioned previously, in the use of $ a(t) \sim a_0 ( t / t_0 )^{2/3}$ in relating 
$G(a)$ in Eq.~(\ref{eq:grun_a}) to $G(t)$ in Eq.~(\ref{eq:grun_t}). 
In general, if $w$ is not small, one should use instead Eq.~(\ref{eq:a_t}) to relate 
the variable $t$ to $a(t)$.
The problem here is that, loosely speaking, for $w \neq 0$ at least
two $w$'s are involved,  $w=0$ (matter) and $w=-1$ ($\lambda$ term).
Unfortunately, this issue complicates considerably the problem of relating 
$\delta G(t)$ to $ \delta G(a) $, and therefore the solution to the 
resulting differential equation for $\delta (a)$.
As a tractable approximation though, one should set instead 
$ a(t) \sim a_0 ( t / t_0 )^{2/3 (1+w)}$,  and then use an ``effective'' value of
$w \approx - 7/9$, which would seem more appropriate for the final target 
value of $\Omega \approx 0.25$.
For this choice one then obtains a significantly reduced power in 
Eq.~(\ref{eq:grun_a}), namely $\gamma_\nu = 3 (1+w)/ 2 \nu = 1 $.
Furthermore, the resulting differential equation for $\delta (a)$, 
Eq.~(\ref{eq:dfq_delta_Gbox_a}), is still relatively easy to solve, by
the same methods used in the previous section.
One now finds 
\beq 
\gamma =  0.5562 - ( 0.92 + 7.70 \, c_h ) \, c_a + O(c_a^2 ) \; .
\label{eq:gamma_ch_1}
\eeq
which should be compared to the previous result of Eq.~(\ref{eq:gamma_ch}).
In particular for the tensor box case one has again $c_h = 7.927 $, which 
can the be used to compare to the previous result of Eq.~(\ref{eq:gamma_relat_ten}).
Thus by reducing the value of $\gamma_\nu$ by about a factor of four, the $c_a$
coefficient in the above expression has been reduced by about a factor of three,
a significant change.

After using this improved value for the power $\gamma_\nu$, the problem
of correcting for relative scales needs to be addressed again,
in light of the corrected estimate for the growth exponent parameter 
of Eq.~(\ref{eq:gamma_ch_1}).
Given this new choice for $\gamma_\nu =1$, on can now consider, for example,
the types of galaxy clusters studied recently in \cite{smi09,vik09,rap09},  which 
typically involve comoving radii of $\sim 8.5 Mpc $ and viral radii of $\sim 1.4 Mpc$.
For these one would obtain an approximate overall scale reduction factor of 
$ (1.4/4890)^{1} \approx 2.9 \times 10^{-4}$.
Note that in these units $(Mpcs)$ the reference scale appearing in $G (\Box ) $ is 
of the order of $\xi \simeq 4890 Mpc$.
This would give for the tensor box ($c_h = 7.927 $ ) correction to the growth index
$\gamma$ in Eq.~(\ref{eq:gamma_ch_1}) the more reasonable order of 
magnitude estimate $ -62. \times 20.6 \times 2.9 \times 10^{-4} \approx - 0.37 $, and
for $\gamma$ itself the reduced value would end up at $\approx 0.19$.
Clearly at this point these should only be considered as rough order of magnitude estimates.

Nevertheless this last case is suggestive of a trend, quite independently of the 
specific value of $c_h$ and therefore of the overall numerical coefficient of
the correction in Eq.~(\ref{eq:gamma_ch_1}): namely that the correction to
the growth index parameter will increase close to linearly 
(for $\gamma_\nu$ close to one, as we have argued) in the size of the cluster.
Consequently one expects that the deviations will increase tenfold in going 
from a cluster size of $1 Mpc$  to one of $10 Mpc$, and a hundredfold in going 
from $1 Mpc$ to $100 Mpc$.

Finally another possible, and ultimately much more conservative, approach would be to take
- at least for the time being - with some caution the rather large value for $c_0$ obtained
from nonperturbative lattice quantum gravity calculations.
One could then use instead the observational bounds on x-ray studies of
large galactic clusters at distance scales of up to about $1.4-8.5 Mpc$ \cite{vik09}, namely
$\gamma= 0.50 \pm 0.08 $, to constrain the value
of the constant $c_a$ {\it at that scale}, giving for example from 
Eq.~(\ref{eq:gamma_ch_1}) the bound $c_a \lsim 8 \times 10^{-4}$ 
in the case of tensor box, and the much less stringent bound
$c_a \lsim O(1) $ for the Newtonian (non-relativistic) case
of Eq.~(\ref{eq:gamma_newt}).

\newpage

\vskip 40pt

\begin{center}
\epsfxsize=10cm
\epsfbox{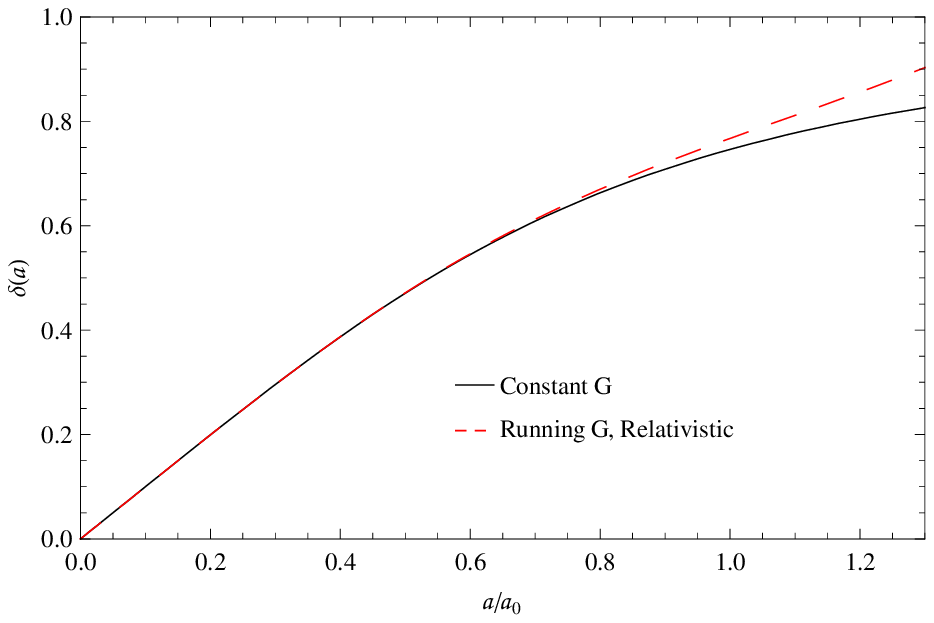}
\end{center}

\noindent{\small Figure 1. Illustration of the matter density contrast $\delta (a)$ as 
a function of the scale factor $a(t)$, in the fully relativistic treatment (tensor box)
and for a given matter fraction $\Omega=0.25$, obtained from the solution of 
the density contrast equation of Eq.~(\ref{eq:dfq_delta_G0}), with $G(a)$ 
given in Eq.~(\ref{eq:grun_a}) with $\gamma_\nu = 9/2$ and for $c_a=0.001$.
In the case of a running $ G(\Box) $, one generally observes a slightly faster growth rate 
for later times, as compared to the solution for the case of constant $G$ and
with the same choice of $\Omega$, described by Eq.~(\ref{eq:dfq_delta_G0_a}).}


\vskip 40pt

\begin{center}
\epsfxsize=10cm
\epsfbox{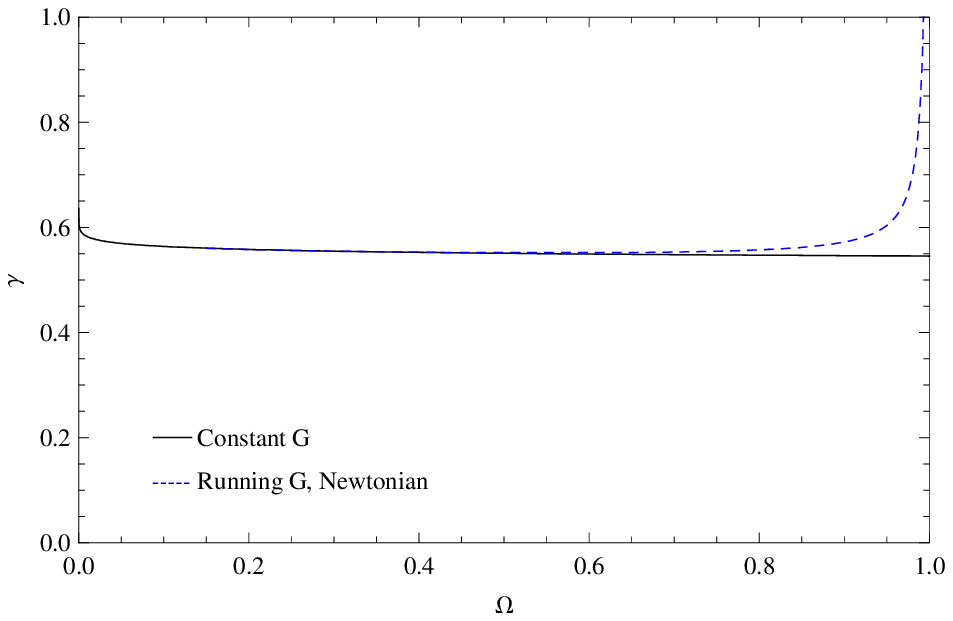}
\end{center}

\noindent{\small Figure 2. Illustration of the growth index parameter $\gamma$ of 
Eq.~(\ref{eq:gamma_def}) as a function of the matter density fraction
$\Omega$, computed in the Newtonian (non-relativistic)
theory with a running $G(a)$ given in Eq.~(\ref{eq:grun_a}), and
obtained by solving Eq.~(\ref{eq:delta_a_eq_newt}), here with 
with $\gamma_\nu = 9/2$ and $c_a=0.01$.
For the specific choice of matter fraction $\Omega = 0.25$, suggested
by $\Lambda CDM$ models, one then obtains the estimates for the
growth index parameter given in Eq.~(\ref{eq:gamma_newt}).}

\newpage

\vskip 40pt

\begin{center}
\epsfxsize=10cm
\epsfbox{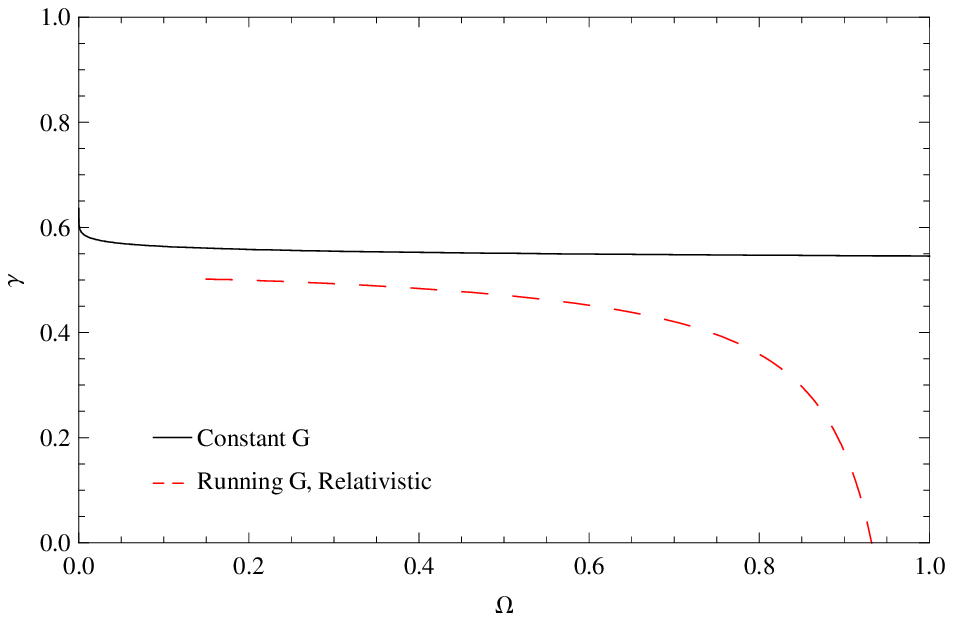}
\end{center}

\noindent{\small Figure 3. Illustration of the growth index parameter $\gamma$ of 
Eq.~(\ref{eq:gamma_def}) as a function of the matter density fraction
$\Omega$, computed in the fully relativistic (tensor box)
theory with a running $G(a)$ as given in Eq.~(\ref{eq:grun_a}), and obtained 
by solving Eq.~(\ref{eq:dfq_delta_Gbox}) with 
with $\gamma_\nu = 9/2$ and $c_a=0.0003$.
For the specific choice of matter fraction $\Omega = 0.25$ one then
obtains the estimates given for the tensor box in 
Eq.~(\ref{eq:gamma_relat_ten}).
Not surprisingly the deviations from the standard result for $\gamma$
become more visible for larger values of $\Omega$.}


\vskip 40pt

\begin{center}
\epsfxsize=10cm
\epsfbox{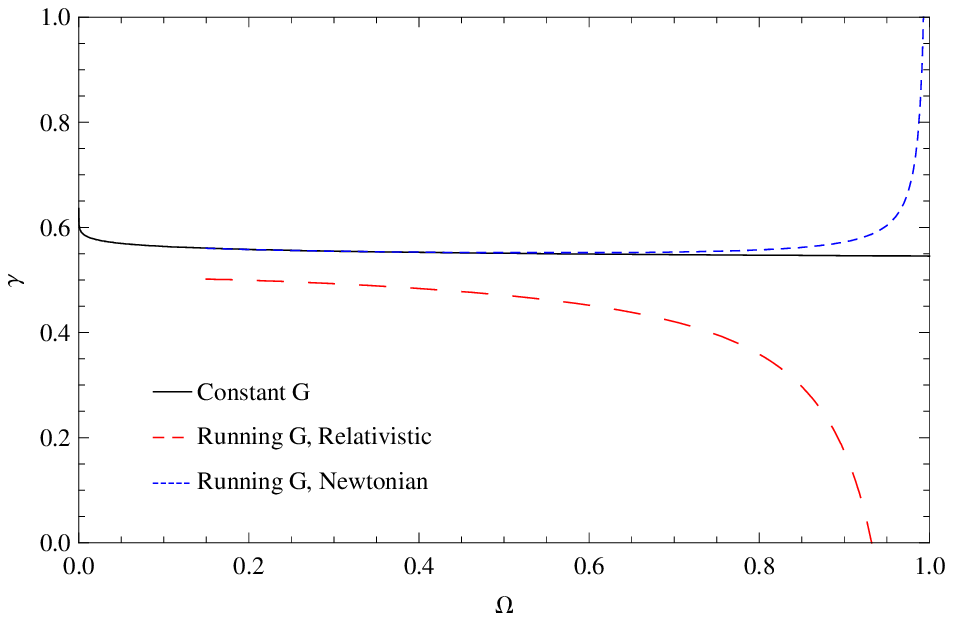}
\end{center}

\noindent{\small Figure 4. Qualitative comparison of the growth index parameters $\gamma$ 
of Eq.~(\ref{eq:gamma_def}) as a function of the matter density fraction
$\Omega$, computed first in the relativistic (tensor box)
theory with a running $G(a)$ and $c_a=0.0003$, then in the Newtonian 
(non-relativistic) treatment also with a running $G(a)$ and $c_a=0.01$, 
both with $\gamma_\nu = 9/2$,
and finally compared to the usual treatment with constant $G$.
In both cases the deviations from the standard result for $\gamma$ are
most visible for larger values of $\Omega$, corresponding to a greater matter fraction.}

\newpage

\vskip 40pt
\subsection{Density Perturbations in the Conformal Newtonian Gauge with ${\bf G(\Box)}$}
\hspace*{\parindent}
\label{sec:conf_newt}

In this section we will outline briefly what other avenues can be pursued to
determine quantitatively and systematically the cosmological effects of a running $G(\Box)$.
The perturbed RW metric is well suited for discussing matter perturbations, but
occasionally one finds it more convenient to use a different metric parametrization,
such as the one derived from the conformal Newtonian (cN) gauge line element
(see for example \cite{mab95,ber01}, and references therein)
\beq
d \tau^2 \; = \; a^2 (t) \left \{  (1+2 \, \psi ) \, dt^2 \; - \; (1-2 \, \phi ) \, \delta_{ij} \, dx^i dx^j \right \}
\label{eq:cn_gauge}
\eeq
with conformal Newtonian potentials $\psi ({\bf x},t)$ and $\phi({\bf x},t)$.
In the simplest framework, the two potentials $\psi$ and $\phi$ give rise separately
to Newton's equation for a point particle, and Poisson's equation, respectively
\beq
\ddot{\bf x} = - {\bf \nabla} \, \psi     \;\;\;\;\;\;\;\;\;\;\;\;   
\nabla^2 \phi = 4 \pi \, G \, a^2 \, \delta \rho \; .
\eeq
In this gauge, and in the absence of a $G(\Box)$, the unperturbed equations are
\bea
\left (  { \dot{a} \over a } \right )^2  & \! = \! &  { 8 \pi \over 3 } \, G \, a^2 \, \bar{\rho}
\nonumber \\
{ d \over d t }  \left (  { \dot{a} \over a } \right  ) & \! = \! &
- { 4 \pi \over 3 } \, G \, a^2 \, ( \bar{\rho} + 3 \, \bar{p} ) \; ,
\label{eq:cn_field_zeroth}
\eea
in the absence of spatial curvature ($k=0$).
In the presence of a running $G$ these again need to be modified, in accordance
with Eqs.~(\ref{eq:fried_rr}),  (\ref{eq:fried_tt}) and  (\ref{eq:grun_t}).
A cosmological constant can be conveniently included in the $\bar{\rho}$ and $\bar{p}$, 
with $\bar{\rho}_\lambda = \lambda/ 8 \pi G = - \bar{p}_\lambda $.
In this gauge scalar perturbations are characterized by Fourier modes 
$\psi({\bf q},t)$ and $\phi ({\bf q},t)$, and the first order Einstein field equations
in the absence of $G(\Box)$ read \cite{mab95}
\bea
k^2 \, \phi \, + \, 3 \, { \dot{a} \over a }  \, \left ( \dot{\phi} \, + \, { \dot{a} \over a }  \, \psi \right )
& \! = \! &
4 \pi \, G \, a^2  \,  \delta T^0_{\;\; 0}
\nonumber \\
k^2  \, \left ( \dot{\phi} \, + \, { \dot{a} \over a }  \, \psi \right )
& \! = \! &
4 \pi \, G \, a^2  \, ( \bar{\rho} + \bar{p} ) \, \theta
\nonumber \\
\ddot{\phi} \, + \, { \dot{a} \over a }  \left ( 2 \dot{\phi} \, + \dot{\psi} \right )
+ \left ( 2 \, { \ddot{a} \over a }  \, - \, { \dot{a}^2 \over a^2 }   \right ) \, \psi
\, + \, { k^2 \over 3 } \, ( \phi \, - \, \psi )
& \! = \! &
{ 4 \pi \over 3 } \, G \, a^2  \,  \delta T^i_{\;\; i}
\nonumber \\
k^2 \, ( \phi \, - \, \psi )
& \! = \! &
12 \pi \, G \, a^2  \,  ( \bar{\rho} + \bar{p} ) \, \sigma \;\;\;\;
\label{eq:cn_field_pert}
\eea
where the perfect fluid energy-momentum tensor is given to linear order in the 
perturbations  
$\delta \rho = \rho - \bar{\rho} $ and $\delta p = p - \bar{p} $
by
\bea
T^0_{\;\; 0}  & \! = \! &  - ( \bar{\rho} \, + \, \delta \rho )
\nonumber \\
T^0_{\;\; i}  & \! = \! &  ( \bar{\rho} \, + \, \bar{p} ) \, v_i  \; = \; - T^i_{\;\; 0}
\nonumber \\
T^i_{\; j}  & \! = \! &  ( \bar{p} \, + \, \delta p ) \, \delta^i_{\; j} \, + \, \Sigma^i_{\; j} 
\;\;\;\;\;\; \Sigma^i_{\; i} = 0
\eea
and one has allowed for an anisotropic shear perturbation $\Sigma^i_{\; j}$ to
the perfect fluid form $T^i_{\; j}$.
 The two quantities $\theta$ and $\sigma$ are commonly defined by 
\beq
( \bar{\rho} \, + \, \bar{p} )  \, \theta \; \equiv \; i \, k^j \, \delta T^0_{\; j}  
\;\;\;\;  
( \bar{\rho} \, + \, \bar{p} )  \, \sigma \; \equiv \; 
- ( \hat{k_i} \hat{k_j} - { 1 \over 3} \delta_{ij} ) \Sigma^i_{\; j}
\eeq
with $\Sigma^i_{\; j} \equiv T^i_{\, j} - \delta^i_{\; j} T^k_{\; k} /3 $ the traceless
component of $T^i_{\; j}$.
For a perfect fluid $\theta$ is the divergence of the fluid velocity, $\theta = i k^j v_j$, 
with $v^j = d x^j / dt $ the small velocity of the fluid.
The field equations imply, by consistency, the covariant energy momentum conservation law
\bea
\dot{\delta} & = & - (1+w) \, (\theta - 3 \dot{\phi} ) - 3 \, { \dot{a} \over a }  \, 
\left ( { \delta p \over \delta \rho } - w \right ) \delta
\nonumber \\
\dot{\theta} & = &  - { \dot{a} \over a }  \, ( 1 - 3 w ) \, \theta - 
{ \dot{w} \over 1+w } \, \theta 
+ { 1 \over 1+ w } \, { \delta p \over \delta \rho }  \, k^2 \delta - k^2 \sigma + k^2 \psi 
\eea
and relate the matter fields $\delta$, $\sigma$ and $\theta$ to the metric perturbations
$\phi$ and $\psi$.
where $\delta $ is the matter density contrast $\delta = \delta \rho / \rho $, and
$w$ is the equation of state parameter $w= p/\rho $.
In General Relativity $\phi = \psi$ as long as there is no anisotropic stress,
but in extended theories of gravity, such as the one described here,
the relation between $\phi$ and $\psi$ can become scale dependent.

In the presence of a $G(\Box)$ the above equations need to be re-derived and amended,
starting from the covariant field equations of Eq.~(\ref{eq:field1}) in the cN gauge
of Eq.~(\ref{eq:cn_gauge}),  with zeroth order modified field equations as
in Eqs.~(\ref{eq:fried_tt}) and (\ref{eq:fried_rr}), using the expansion
for $G(\Box)$ given in Eq.~(\ref{eq:gbox_h}), but now in terms of the new cN
gauge potentials $\phi$ and $\psi$.
One key question is then the nature of the vacuum-polarization induced 
anisotropic shear perturbation correction $\Sigma^i_{\; j}$ 
appearing in the covariant effective field equations analogous to 
Eqs.~(\ref{eq:cn_field_pert}), but derived with a $G(\Box)$.
In particular  one would expect the quantum correction to the energy momentum
tensor appearing on the r.h.s. of Eq.~(\ref{eq:field1}) to contribute new terms
to the last of Eqs.~(\ref{eq:cn_field_pert}),
which could then account for a non-zero stress $\sigma$, and thus for a small 
deviation from the classical GR result for a perfect fluid, $\phi = \psi$.
Naively one would expect $ \psi / \phi = 1 + O( \delta G / G_0) $.
An explicit calculation with $G(\Box)$ \cite{rei10} gives
\beq
{ \psi \over \phi }  \, = \,
1 + \left ( 1 - { 1 \over 2 \nu \, (1 + w) } \right ) \, 3 \, w_{vac} \, { \delta \, G(t) \over G_0 }
\, = \,
1 + \left ( 1 - { 1 \over 2 \nu } \right ) \, { \delta \, G(t) \over G_0 } \; 
\eeq
for $w=0$ and $w_{vac} = \third $.
It is often customary (see e.g. \cite{mab95,ber01,ame07,dan09}) to parametrize 
deviations from General Relativity in terms of a slip
function $\Sigma$ and of the growth rate parameter $\gamma$ introduced
previously.
These two quantities are defined by
\beq
\nabla^2 ( \phi + \psi ) = 3 \, \Sigma \, \Omega \, H^2 \, \delta 
\;\;\;\;\;\;\;\;  \gamma  = { \log f \over \log \Omega } 
\eeq 
with $\delta$ the density contrast and $f$ the density contrast exponent.
Occasionally the parameter $\eta = \psi / \phi -1 $ is introduced as well.
In classical General Relativity $\psi / \phi = 1$, $\eta =0 $,
$\Sigma=1$ and then the growth exponent $\gamma \approx 0.55$ for $\Omega \approx 0.25$.
The calculations presented in the previous sections have already suggested to 
some extent what changes to expect for the exponent $\gamma$, which then 
leaves the problem of determining the structure of the $\Sigma$ correction.
In addition, the Newtonian (non-relativistic) calculation of Appendix A has determined,
from the form of the modified Poisson equation, one of the relevant equations,
namely the one for the potential $\phi$.
We plan to discuss these interesting questions in a future publication \cite{rei10}.


\vskip 40pt
\newsection{Conclusions}
\hspace*{\parindent}

In this paper we have attempted to systematically analyze the effects on matter 
density perturbations of a running $G(\Box)$ appearing in the original 
effective, non-local covariant field equations of Eq.~(\ref{eq:field1}).
The specific form of $G(\Box)$ in Eq.~(\ref{eq:grun_box_0}) is inspired by the 
non-perturbative treatment of covariant path integral quantum gravity, and follows 
from  the existence of a non-trivial fixed point in $G$ of the renormalization group in
four dimensions.
The resulting effective field equations are manifestly covariant, and in principle
besides the genuinely non-perturbative scale $\xi$
there are no adjustable parameters, since the coefficients ($c_0$) and scaling
dimensions ($\nu$) entering $G(\Box)$ are, again in principle, calculable by systematic 
field theory and lattice methods (\cite{book} and references therein).

The present work can be viewed in broad terms as consisting of two parts.
In the first part we have systematically developed the general formalism necessary to deal 
with small matter density fluctuations in the presence of a running gravitational coupling $G(\Box)$.
Most, if not all, of the results in the first part have been formulated in a way that 
assumes as little as possible about specific aspects related to how exactly
$G$ does run with scale.
Indeed many of the equations we have obtained are not restricted to $\nu = \third $,
and are found to be valid for a wide range of 
powers $\nu$ and coefficients $c_0$  appearing for example in the original expression 
for $G(\Box)$ as given in Eq.~(\ref{eq:grun_box}).
Furthermore, the zeroth order (in the fluctuations) results of \cite{hw05}, on which the present
work builds up, do not rely on any specific value for these parameters either, since the
expressions obtained there follow from general properties of the covariant d'Alembertian 
and its powers, as they appear in $G(\Box)$.
In particular the flow in the vicinity of the ultraviolet fixed point could in
principle allow for $c_0$ being either negative (gravitational screening)
or positive (gravitational anti-screening), and both cases could in principle
be described by the results obtained above, for example for the growth index $f$
and the growth index parameter $\gamma$.
It is only the latter option though that is favored by studies of non-perturbative Euclidean
lattice gravity (the weak coupling phase is unstable and found to describe a collapsed 
degenerate two-dimensional spacetime), hence the choice here to discuss primarily 
this last case.
But in principle the fact remains that the sign of $c_0$  will ultimately determine
the direction of the corrections given above,  which could 
eventually become constrained by observation.
In the end the only result that is extensively used in the first part is the result of \cite{hw05}
that $w_{vac}=\third$, apart from the fact that we choose to restrict our attention from
the very beginning primarily to the non-relativistic matter case $w=0$, and to the large 
wavelength limit ${\bf q} \rightarrow 0$.
Later on it was found that for sufficiently slowly varying backgrounds the result 
$w_{vac}=\third$ is preserved also to first order in the perturbations, which seems
to suggest some level of consistency in the treatment of the field perturbations.

In spite of the non-locality of the original effective field equations in Eq.~(\ref{eq:field1}),
one finds quite in general that small perturbations can be treated, in a first approximations, in terms
of local terms, described by quantities $\rho_{vac}$ and $p_{vac}$ as they appear
in the effective description of $T_{\mu\nu}^{vac}$ in terms of a perfect fluid.
The latter should then be regarded as the leading term in a derivative expansion of 
the non-local contribution to the effective field equations, as they apply here to 
the rather specific case of the $FRW$ background.
Under the physically motivated assumption of a comparatively slowly varying (both in space and time) 
background, it is then possible to obtain a complete and consistent set of effective 
field equations, describing small perturbations for the metric trace and matter modes
(Eqs.~(\ref{eq:encons_fluc}), (\ref{eq:field_fluc_tt}), (\ref{eq:field_fluc_ii})
and (\ref{eq:field_fluc_h})).
From these a single equation for the matter density contrast is eventually obtained,
Eq.~(\ref{eq:dfq_delta_Gbox}), which is the main result of this work.
The only input needed in this last equation is $\delta G(t)$, the zeroth order 
(in the fluctuations) running of $G$ as written in Eq.~(\ref{eq:grun_t}), with 
given more or less known parameters $\nu$ and $c_t$.
The corresponding result in the Newtonian (non-relativistic) treatment is
obtained in Appendix A, leading to Eq.~(\ref{eq:dfq_delta_newtonian_leading_Gbox}).

The next step was a translation of the equation for the density contrast $\delta (t)$
into the corresponding equation for $\delta (a)$, involving a related running coupling $G(a)$,
instead of the original $G(t)$.
Since in general the transformation from one variable to the other is not entirely trivial,
some simplification had to be assumed, i.e. that the quantum correction in $G(a)$ can
be written as a power, with an exponent $\gamma_\nu$, a choice that could
in the future be relaxed as part of a broader more systematic investigation.
Subsequently a solution for the differential equation for $\delta (a)$ was obtained, 
leading to expressions for the growth index $f(a)$ and for the growth index parameter $\gamma$.
A number of general features can be observed, the first one being the fact that
generally the correction to the growth index parameter $\gamma$ is found to
be negative, indicating a less steep rise of $f$ with $\Omega$.

The second part of the paper describes a number of attempts to provide a 
semi-quantitative estimate for the corrections obtained, in order to see whether 
these corrections could be related in some way to current astrophysical observations.
In order to do so, one needs to adapt the theoretical calculation for the growth index 
parameter $\gamma$ to the kind of observational data available from the study of 
large galactic clusters.
This requires, as expected, a careful consideration of the relative length scales that come into play.
On the one hand, one length scale is given by the size of the largest clusters reached by 
observation, typically of the order of a few $Mpc$s.
On the other hand it should involve the absolute reference scale given by
$\xi = \sqrt{ 3/ \lambda } \simeq 4890 Mpc$.
The comparison between theory and observation would then seem straightforward,
were it not for the fact that this ratio generally comes in to a certain power, whose 
detailed knowledge is necessary in order to eventually reduce the quantitative uncertainties.
Eventually these could be bracketed by a more systematic study of
the solutions to the $ \delta (a) $ equation, and the corresponding growth exponents 
$\gamma$.
We are referring here in particular to a study of the sensitivity of the results to the specific 
choices of the exponent $\gamma_\nu$, appearing in $\delta G(a)$ and determined
in part by the relationship between the variables $t$ and $a(t)$, which we discussed earlier.
In addition, there is still perhaps a certain level of uncertainty in the actual coefficients
$c_0$ and $c_t$ entering the theoretical predictions, which we have also 
described above in some detail.
The latter could be reduced further by improved non-perturbative lattice computations.
Nevertheless, the value of the present calculations lies in our opinion in the fact
that so far a discernible trend seems to emerge from the results.
The trend we have found seems to suggest that the correction to the growth exponent 
$\gamma$ is initially rather small for small clusters, negative in sign,
and then slowly increasing in magnitude, close to linearly with scale.

It is clear that the effects discussed in this paper are only relevant for very large scales,
much bigger than those usually considered, and well constrained, by laboratory, solar
or galactic dynamics tests \cite{dam06,uza02,uza09,ade03}.
Furthermore the effects we have described here are quite different from what one 
would expect in $f(R)$ theories \cite{cap09,car04}, which also tend to predict
some level of deviation from classical GR in the growth exponents \cite{mot10,zha05,gan08}.
Future more accurate astrophysical observations might make it possible to see the difference
in the predictions of various models \cite{ame07,dan09,gol08,ber93,rob07}.


\vspace{20pt}

{\bf Acknowledgements}

One of the authors (HWH) wishes to thank Thibault Damour and Gabriele Veneziano
for inspiration and discussions leading to the present work, and
Alexey Vikhlinin for correspondence regarding astrophysical measurements 
of structure growth indices.
He also wishes to thank Hermann Nicolai and the
Max Planck Institut f\" ur Gravitationsphysik (Albert-Einstein-Institut)
in Potsdam for warm hospitality. 
Thanks also go to  Salvatore Capozziello and Paolo Serra for providing us 
with a number of useful references.
The work of HWH was supported in part by the Max 
Planck Gesellschaft zur F\" orderung der Wissenschaften, and
by the University of California.
The work of RT was supported in part by a DoE GAANN student fellowship.

\newpage

\appendix

\section*{Appendix}

\vskip 40pt
\newsection{Non-Relativistic (Newtonian) Treatment of Matter Density Perturbations}
\hspace*{\parindent}
\label{sec:newt}

In this section we discuss the Newtonian theory of small matter fluctuations, first
by recalling the relevant equations in the usual treatment, and then by presenting
what changes need to be implemented in order to account for the running of $G$.
Later these equations will be solved, so that a comparison can be made with the 
results in the absence of a running $G$.

When discussing a nonrelativistic Hubble flow it is customary to define coordinates
in the following way
\beq
{\bf x} = { {\bf r} \over a (t)} \;\;\;\;\;\;\;\;\;\;\;\;\;\;\;\;\;\;
 {\bf v} = { d {\bf r} \over d t} = { \dot{a} \over a } \; {\bf r} 
\label{eq:hubble_flow}
\eeq
where $ {\bf x} $ is attached to the comoving frame, while $ {\bf r} $ is the flat Minkowski
space coordinate, such that in the comoving frame $ {\bf x} $ one has, by construction,
$ d {\bf x} / d t = 0 $.

In the following some simplification will arise due to the fact that we shall consider
a non-relativistic fluid with the negligible pressure, $p \simeq 0$ or $w=0$.
The relevant equations are then the continuity equation, the Euler equation and the gravitational
field equations.
These will be listed below to zeroth and first order in the matter density ($\rho$), pressure ($p$), velocity field (${\bf v}$) and gravitational field ${\bf g}$.

\vskip 40pt
\subsection{Newtonian Treatment {\it\bf Without} the Running of ${\bf G}$}
\hspace*{\parindent}
\label{sec:newtonian}

After decomposing the fields into a background and a fluctuation contribution, 
$\rho = \bar{\rho}+\delta \rho$, $p = \bar{p}+\delta p $, and ${\bf v} = \bar{\bf v}+\delta {\bf v}$, 
one obtains from the continuity equation, to zeroth and first order respectively,
\beq
\dot{\bar{\rho}} + {\bf \nabla} \cdot \left ( \bar{\rho} \, {\bf v} \right ) = 0  
\;\;\;\;\;\;\;\;\;\;\;\;\;\;\;\;\;\;\;\;\;\;
\dot{\delta \rho} + 3 \, {\dot{a} \over a} \, \delta \rho + {\dot{a} \over a } \, 
\left ( {\bf r} \cdot {\bf \nabla}\right ) \, 
\delta \rho + \bar{\rho} \, {\bf \nabla} \cdot \delta {\bf v} = 0 \; .
\label{eq:newt_cont}
\eeq
When the effect of the Hubble flow is included, {\it i.e.,} Eq.~(\ref{eq:hubble_flow}), the above 
zeroth order equation reduces to
\beq
\dot{\bar{\rho}} (t) + 3 \, {\dot{a} (t) \over a (t)}  \, \bar{\rho}(t) \, = 0
\label{eq:cont_frw}
\eeq
with solution $ \bar{\rho} (t) = \bar{\rho}_0 \, \left ( a_{0} /  a (t) \right )^3 $,
where $\bar{\rho}_0 $ and $ a_{0} $ are the two integration constants corresponding 
to the present matter density and to the present scale factor (usually taken to be $a_{0} = 1$).
We note here that Eq.~(\ref{eq:cont_frw}), and hence 
Eq.~(\ref{eq:rho_fried}), will continue to hold for a running $ G $, 
as these equations are derived from the
kinematics and the continuity equations in the RW background metric given in
Eq.~(\ref{eq:newt_cont}), whose is not affected by the running of $G \rightarrow G(\Box) $.

To zeroth and first order in the fluctuations the Euler equations for a fluid in the RW background
are given respectively by
\beq
\dot{{\bf v}} + \left ({\bf v} \cdot {\bf \nabla} \right ) \, {\bf v} =  {\bf g}
\;\;\;\;\;\;\;\;\;\;\;\;\;\;\;\;\;\;\;\;\;\;
 \dot{\delta {\bf v}} + {\dot{a} \over a} \, \delta {\bf v} + {\dot{a} \over a} \,
\left ( {\bf r} \cdot {\bf \nabla} \right ) \, \delta {\bf v} = - {1\over \bar{\rho}} \,
{\bf \nabla} \, \delta p + \delta {\bf g} \; .
\label{eq:newt_euler}
\eeq
Finally the gravitational field equations are given to zeroth and first order in the 
fluctuations by 
\beq
{\bf \nabla} \times {\bf g} = 0 
\;\;\;\;\;\;\;\;\;\;\;\;\;\;\;\;\;\;\;\;\;\;\;\;\;\;\;\;\;\;\;\;
{\bf \nabla} \cdot {\bf g} = - \, 4 \pi \, G_0 \, \bar{\rho}
\label{eq:newt_grav}
\eeq
\beq
{\bf \nabla} \times \delta {\bf g} = 0
\;\;\;\;\;\;\;\;\;\;\;\;\;\;\;\;\;\;\;\;\;\;\;\;\;\;\;\;\;\;\;\;
{\bf \nabla} \cdot \delta {\bf g} = - \, 4 \pi \, G_0 \, \delta \rho \;
\label{eq:newt_grav1}
\eeq
incorporating Gauss' law and the constraint that the gravitational fields are longitudinal.
Only the last set of equations contain the gravitational constant $G$.
Hence, in the framework of the Newtonian treatment, the modification of 
a running $G \rightarrow G(\Box) $ only affects the gravitational Poisson equation.

It is customary at this stage to introduce Fourier components of the fluctuations,
and write 
\beq
\delta \rho ({\bf r}, t) = \delta \rho_{\bf q} (t) \, 
\exp \left [ { i\, {\bf r} \cdot {\bf q} \over a(t)} \right ]
\label{eq:fouriertrans}
\eeq
and similarly for $\delta {\bf v}$, $\delta {\bf g}$ , and possibly $\delta p $.
For an adiabatic fluctuation one can also set $ \delta p = v_s^2 \, \delta \rho$,
with $v_s$ the speed of sound.

Then to first order in the fluctuations the continuity equation,  Euler equation  and the
gravitational field equations take on the form, for each mode ${\bf q}$,
\beq
\dot{\delta \rho}_{\bf q} (t) + 3 \, {\dot{a} (t) \over a(t) } \, \delta \rho_{\bf q} (t)  +
{i \, {\bf q} \cdot \delta {\bf v}_{\bf q} (t) \over a(t) } \, \bar{\rho} (t) = 0
\label{eq:cont_fluc}
\eeq

\beq
\dot{\delta {\bf v}}_{\bf q} (t) + {\dot{a}(t) \over a(t)} \, \delta {\bf v}_{\bf q} (t) =
-  \, {i \, {\bf q} \over a(t)} \, {v_s^2 \over \bar{\rho} (t) } \, \delta \rho_{\bf q} (t) +
\delta {\bf g}_{\bf q} (t)
\label{eq:euler_fluc}
\eeq

\beq
\delta {\bf g}_{\bf q} (t) = { 4 \pi i \,  {\bf q} \over {\bf q}^2 }
\,  a (t) \,  G_0 \, \delta \rho_{\bf q} (t) \; .
\label{eq:g1_G0}
\eeq
Subsequent elimination of the gravitational and velocity fields then leads to a single second order
differential equation for the matter density contrast 
$\delta_{\bf q} (t) \equiv \delta \rho_{\bf q} (t) / \bar{\rho} (t) $ describing the
physics of compressional modes:
\beq
\ddot{\delta}_{\bf q} (t) + 2 \, {\dot{a} (t) \over a (t) } \, \dot{\delta}_{\bf q} (t)
+ \left ( {v_s^2 \, {\bf q}^2 \over a (t)^2 } - 4 \pi \, G_0 \, \bar{\rho} (t) \right ) \, 
\delta_{\bf q} (t) = 0 \; .
\label{eq:dfq_delta_newtonian_exact_G0}
\eeq
In the limit of very long wavelength fluctuations, 
$ {\bf q} \rightarrow 0 $, the above equation simplifies to
\beq
\ddot{\delta}(t) + 2 \, {\dot{a} (t) \over a (t)} \, \dot{\delta}(t) 
-  4 \pi \, G_0 \, \bar{\rho} (t) \, \delta(t) = 0 \; .
\label{eq:dfq_delta_newtonian_leading_G0}
\eeq
A solution can then be found, using $\bar{\rho} (t)=1/ 6 \pi G t^2 $ and
$ \dot{a} (t) / a (t) \equiv H(t)=2/3 t $, such that the general form
for $\delta (t)$ is given by a linear combination of either $\sim t^{2/3}$ or $\sim t^{-1}$.
The latter corresponds to a decaying (as opposed to growing) solution and
is usually discarded, giving finally the standard Newtonian result $\delta(a) \propto a$.
We note here that the above non-relativistic equation and solution applies to the case
of non-relativistic matter only; in particular it excludes the presence of a cosmological constant.

\vskip 40pt
\subsection{Newtonian Treatment with Running ${\bf G(\Box)}$}
\hspace*{\parindent}
\label{sec:newtonian-run}

The next step is a modification of the non-relativistic equations in Eqs.~(\ref{eq:newt_cont}),  (\ref{eq:newt_euler}),  (\ref{eq:newt_grav})  and (\ref{eq:newt_grav1}) 
to incorporate a suitable running of $G$.
Since only the latter set of equations, Eqs.~(\ref{eq:newt_grav}) and (\ref{eq:newt_grav1}), 
contain $G$ it is only these that need to be suitably modified.
In the presence of a scale-dependent coupling one has
\beq
\delta {\bf g} = - {\bf \nabla} \, \delta \phi
\eeq
with the perturbing potential $\delta \phi$ given by a solution to Poisson's equation
\beq
{\bf \nabla}^2 \delta \phi ({\bf r}, t) = - {\bf \nabla} \cdot \delta {\bf g} ({\bf r}, t) = 4
\pi G (\Box)\, \delta \rho ({\bf r}, t)
\eeq
and $G(\Box)$ given in Eq.~(\ref{eq:grun_box}).
Following Eq.~(\ref{eq:fouriertrans}), as it applies here to $\delta {\bf g}$ and $\delta \rho$, we
Fourier transform the spatial components of the above Poisson equation, which
requires the Fourier transform of $ G(\Box) $ as obtained from Eq.~(\ref{eq:grun_box}), namely
\beq
G( {\bf q}^2, \partial_t^2) = G_0 \, \left \{  1 + c_0 \, {\xi^{-1/\nu} \over 
\left [ - \partial_t^2 -  {\bf q}^2 / a^2(t)  \right ]^{1 / 2 \nu} } + \dots
\right \} \; .
\eeq
As a result the gravitational field perturbation is of the form
\beq
\delta {\bf g}_{\bf q} (t) = { 4 \pi  \, i \, {\bf q} \over {\bf q}^{\,2} } \, a(t) \, \cdot \,
\exp \left [
{- \, i \, {\bf r} \cdot {\bf q} \over a(t) }
\right ] 
\, G ({\bf q}^{\,2}, \partial_t^2) \, 
\left ( 
\delta \rho_{\bf q} (t) \,  \exp 
\left [ { i \, {\bf r} \cdot {\bf q} \over a(t) }  \right ] 
\right ) \; .
\label{eq:g1_exact_Gbox}
\eeq
Since we are mainly interested in the long wavelength limit,
it suffices here to evaluate the above expression in the limit 
$ {\bf q} \rightarrow 0 $,
\bea
\delta {\bf g}_{\bf q} (t) & = & { 4 \pi \, i \, {\bf q} \over {\bf q}^{\,2} } \, a(t) \, 
\left [ 1 - {i \, {\bf r} \cdot {\bf q} \over a(t) } + \dots \right ] \,
G ({\bf q}^{\,2}, \partial_t^2) \, \left ( \delta \rho_{\bf q} (t) \, 
\left [ 1 + {i \, {\bf r} \cdot {\bf q} \over a(t)} + \dots \right ] \right ) 
\nonumber \\
& \simeq & { 4 \pi \, i \, {\bf q} \over {\bf q}^{\,2} } \, a(t) \, 
\left [ G
({\bf q}^{\,2}, \partial_t^2) \, \delta \rho_{\bf q} (t) 
- {i \, {\bf r} \cdot {\bf q} \over a(t)} \, 
G ({\bf q}^{\,2}, \partial_t^2) \, \delta \rho_{\bf q} (t)  
+ G ({\bf q}^{\,2}, \partial_t^2)\, \delta \rho_{\bf q} (t) \, 
{i \, {\bf r} \cdot {\bf q} \over a(t)}  + \dots \right ] \; ,
\nonumber \\
\eea
and for $ {\bf q} = 0 $ only the first term survives.
Furthermore,  when $ G (\Box) = G({\bf q}^2, \partial_t^2) $ acts on a function of $ t $ 
which we will assume here is of the form of a power
({\it e.g.}, $t^{\alpha}$,  with the power $\alpha$ a number of order one) 
one obtains
\beq
G({\bf q}^{\,2}, \partial_t^2) \cdot t^{\alpha} \; \rightarrow \; G(t) \cdot  t^{\alpha} \; .
\eeq
Here the running coupling $G(t)$ is given by the expression in Eq.~(\ref{eq:grun_t}),
with $t_0 \equiv \xi$, and the coefficient
\beq
c_t \; = \; \left \vert { \Gamma ( 1 + \alpha ) \over \Gamma ( 1 + \alpha + 1 / \nu ) } \right \vert \, c_0 \, .
\eeq
Thus for example for $\alpha =-4/3 $ (the standard Newtonian result for matter density
perturbations) one has $c_t = (27/10) \, c_0 $;
in the following it will be safe to assume that the coefficient $c_t$ 
in Eq.~(\ref{eq:grun_t}) is a number of the same order 
of magnitude as the original $c_0$ in Eq.~(\ref{eq:grun_box}).

Consequently, when acting on a density perturbation $\delta \rho_{\bf q} (t) $ in the form
of a power law in $ t $, to leading order in $ {\bf q} $ one obtains simply
\beq
\delta {\bf g}_{\bf q} (t) = {4 \pi \, i  \, {\bf q} \over {\bf q}^{\,2} } \; a(t) \,
G_0 \, \left [ 1 + c_t \, \left (  {t \over t_0} \right )^{1 / \nu} + \dots \right ]
\, \delta \rho_{\bf q} (t) \; .
\label{eq:g1_leading_Gbox}
\eeq
This last result can be compared with Eq.~(\ref{eq:g1_G0}) for the case of a constant $G$.

As stated previously, the continuity equation for the fluctuations, Eq.~(\ref{eq:cont_fluc}), 
and the corresponding Euler equation for the fluctuations,  Eq.~(\ref{eq:euler_fluc}), are not
modified by the presence of a running  $G(\Box)$, as given in 
Eqs.~(\ref{eq:g1_exact_Gbox}) and (\ref{eq:g1_leading_Gbox}).
To solve the resulting equations of motion for the fluctuations, it is now customary to decompose 
the velocity perturbation $ \delta {\bf v} $ into parts perpendicular and parallel to $ {\bf q} $
\beq
\delta {\bf v}_{\bf q} (t) = 
\delta {\bf v}_{ {\bf q} \, \perp } (t) + i \, {\bf q} \, \epsilon_{\bf q} (t)
\eeq
with
\beq
{\bf q} \cdot \delta {\bf v}_{ {\bf q} \, \perp } \, = \, 0 
\;\;\;\;\;\;\;\;\;\;\;\;\;\;\;\;\;\;\;\;\;
\epsilon_{\bf q} \, \equiv \, 
- \, { i \, {\bf q} \cdot \delta {\bf v}_{\bf q}  \over {\bf q}^2}  \; .
\eeq
The fractional change in the matter density $\delta$ is then defined as
\beq
\delta_{\bf q} (t) \equiv { \delta \rho_{\bf q} (t) \over \bar{\rho} (t) }  \; .
\eeq
With the above decomposition of the velocity field $\delta {\bf v} $ and the expression 
for the density contrast $\delta$ inserted into the first order continuity equation, 
Eq.~(\ref{eq:cont_fluc}), one obtains the unmodified result
\beq
\dot{\delta}_{\bf q}(t) = { {\bf q}^2 \over a(t) } \, \epsilon_{\bf q} (t) \; ,
\label{eq:cont_fluc_decompose}
\eeq
so that there is no change in the relationship between $\delta$ and $\epsilon$ when 
$ G \rightarrow G(\Box) $.
In turn the Euler equation for the fluctuation, Eq.~(\ref{eq:euler_fluc}), now becomes
the two sets of equations
\bea
&& Re \, : \;\;    \dot{ \delta {\bf v}}_{ {\bf q} \, \perp} (t) + {\dot{a} \over a} \,
\delta {\bf v}_{ {\bf q} \, \perp} (t)= 0 \nonumber \\
&& Im \, : \;\;    i \, {\bf q} \, \dot{\epsilon}_{\bf q} (t) + {\dot{a} \over a} \, i
\, {\bf q} \, \epsilon_{\bf q} (t) = - \, {i \, {\bf q} \over a}  \, v_s^2\,
\delta_{\bf q} (t) + \delta {\bf g}_{\bf q}
\label{eq:euler_fluc_decompose}
\eea
with the gravitational field fluctuation $\delta {\bf g}_{\bf q} $ now given by the expression
in Eq.~(\ref{eq:g1_exact_Gbox}).
From the real part (corresponding to rotational modes) one concludes
\beq
\delta {\bf v}_{ {\bf q} \, \perp} \propto a^{-1} (t) \; ,
\eeq
which is of the same form as in the case of a constant $G$.
From the imaginary part (corresponding to compressional modes) in 
Eq.~(\ref{eq:euler_fluc_decompose}) one obtains, using Eq.~(\ref{eq:cont_fluc_decompose}),
\beq
\ddot{\delta}_{{\bf q}} (t) 
+ 2 \, {\dot{a} \over a} \, \dot{\delta}_{\bf q} (t)
+ { {\bf q}^2 \over a^2} \, v_s^2 \, \delta_{{\bf q}} (t)
- 4 \pi \, 
\exp \left [ {- \, i {\bf r} \cdot {\bf q} \over a(t)}\right ] 
\, G ({\bf q}^{2}, \partial_t^2) \, 
\left ( \,  
\exp \left [ {i {\bf r} \cdot {\bf q} \over a(t)} \right ] \, 
\bar{\rho}(t) \, \delta_{\bf q}(t) 
\right ) = 0 \; .
\eeq
The latter can be recast into the slightly simpler form 
\beq
\ddot{\delta}_{{\bf q}} (t) 
+ 2 \, {\dot{a} \over a} \, \dot{\delta}_{{\bf q}} (t) 
+ \left ( 
{ {\bf q}^2 \over a^2} \, v_s^2  - 4 \pi \, {\cal G} \, ({\bf q}^{\,2},
\partial_t^2) 
\right ) \, \delta_{\bf q}(t) = 0
\eeq
by defining  a modified source term
\beq
{\cal G} ({\bf q}^{\,2}, \partial_t^2) \equiv
{1 \over \delta_{\bf q}(t) } \,
\left \{ 
\exp \left [ {- \, i {\bf r} \cdot {\bf q} \over a(t)} \right ] \, 
G ({\bf q}^{\,2}, \partial_t^2) \, 
\left ( \,  
\exp \left [ {i {\bf r} \cdot {\bf q} \over a(t)} \right ] \, 
\bar{\rho}(t) \, \delta_{\bf q}(t) \right ) 
\right \} \; .
\eeq
In the limit $ {\bf q} \rightarrow 0 $ one obtains immediately
\beq
\ddot{\delta} (t) 
+ 2 \, {\dot{a} \over a} \, \dot{\delta} (t) 
- 4 \pi \, G (t) \, \bar{\rho} (t) \, \delta(t) = 0 \; .
\label{eq:dfq_delta_newtonian_leading_Gbox}
\eeq
The last two equations can now be compared with the corresponding results
for a constant $G$, given in Eqs.~(\ref{eq:dfq_delta_newtonian_exact_G0}) and
(\ref{eq:dfq_delta_newtonian_leading_G0}).

\vskip 40pt
\subsection{Computation of the Non-Relativistic (Newtonian) Growth Index with ${\bf G(\Box)}$}
\hspace*{\parindent}
\label{sec:newtonian-growth}

The next step requires a solution of the differential equation  for the density perturbations 
$\delta_{{\bf q}} (t) $, in the Newtonian approximation and in the limit  ${\bf q} \rightarrow 0 $, 
as in Eq.~(\ref{eq:dfq_delta_newtonian_leading_Gbox}).
It is convenient and customary at this point to change variables from $t$ to the scale factor $a(t)$,
so that $\delta_{\bf q} (t) \rightarrow \delta_{\bf q} (a) = \tilde{\delta}_{\bf q} \cdot \delta(a) $.
From Eq.~(\ref{eq:a_ttoa}) one has
\bea
\dot{ \delta }(t)& = & a \, H (a) \, {\partial \, \delta (a) \over \partial a} 
\nonumber \\
\ddot{ \delta } (t)& = & {a}^2 \,  H^2 (a) \left [ {\partial \ln H (a) \over \partial a} 
+ {1 \over a} \right ] \,{\partial \, \delta (a) \over \partial a} 
+ {a}^2 \,  H^2 (a) \, {{\partial}^2 \delta (a) \over \partial {a}^2} \; .
\label{eq:d_ttoa}
\eea
Here $H(a)$ is defined as the Hubble ``constant" $ H (a) \equiv \dot{a} (t) / a (t)$,
as it appears in the equations of motion for a background FLRW geometry
\beq
H (a) = \sqrt{ {8 \pi \over 3} \,  G(a) \, \bar{\rho} (a) + {\lambda \over 3} } \; ,
\label{eq:h_frw}
\eeq
but with a running Newton's constant $G(a)$ (see Eq.~(\ref{eq:grun_t}))
\beq
G(a) = G_0 \left [ 1 + { \delta G(a) \over G_0 } \right ]
= G_0 \left [ 1 + c_a \, \left ( {a \over a_0 } \right )^{\gamma_\nu} + \dots \right ] \; .
\label{eq:grun_newt_a}
\eeq
Here the index is $ \gamma_\nu = 3/2 \nu $, since from Eq.~(\ref{eq:grun_t}) 
one has for non relativistic matter $a(t)/a_0 \approx (t/t_0)^{2/3}$.
In the above expression $c_a \approx c_t$ if $a_0$ is identified with a scale factor 
corresponding to a universe of size $\xi$; to a good approximation this corresponds
to the universe ``today'',  with the relative scale factor customarily normalized to $a/a_0=1$.
As a consequence, the constant $c_a$ in Eq.~(\ref{eq:grun_newt_a}) can be taken to be of the 
same order as the constant $c_0$  appearing in the original expressions for $G(\Box)$ in
Eqs.~(\ref{eq:grun_k}) and (\ref{eq:grun_box}).
Note also that, by the use of Eq.~(\ref{eq:h_frw}) for the scale factor,
we have allowed for a non-vanishing cosmological constant in our otherwise 
Newtonian (non-relativistic) treatment. 

After these substitutions one finally obtains the differential equation for the matter density contrast, Eq.~(\ref{eq:dfq_delta_newtonian_leading_Gbox}), in the variable $a(t)$
\beq
{d^2 \delta (a) \over d a^2 } + \left ( { d \ln H (a) \over d a} + {3 \over a} \right ) \, {d
\delta (a) \over d a} - { 4 \pi G(a) \, \bar{\rho}(a) \over a^2 \, H^2 (a) } \, \delta (a) = 0 \; .
\eeq
Note that in order to compute the leading, in $ \delta G (a) / G_0 $, correction to the density 
contrast $\delta (a)$, one only needs 
$\bar{\rho}(a) $ to lowest order as given in Eq.~(\ref{eq:rho_fried}), and $H(a)$ as
given in Eq.~(\ref{eq:h_frw}).

With the aid of the parameter $ \theta $ (see Eq.~(\ref{eq:theta_def}))
\beq
\theta  \equiv  { 1 - \Omega \over \Omega } 
\eeq
where $\Omega$ is the matter density fraction and $1-\Omega$ the 
cosmological constant fraction as measured today,
one obtains the following differential equation for the density contrast $ \delta(a) $
\beq
{\partial^2 \delta \over \partial a^2 }  \! + \! 
{ 3 \left (1 \! + \! 2 \, a^3 \, \theta \right ) \over 
2 \, a \left ( 1 \! + \! a^3 \, \theta \right ) } 
\left (
1 \! + \! c_a {  \gamma_\nu \, a^{ \gamma_\nu} 
\! + \! \left ( \third \gamma_\nu - 1 \right )  a^{3\! + \! \gamma_\nu} \theta  
\over  \left (1 \! + \! a^3 \, \theta \right ) \, \left (1 \! + \! 2 a^3 \, \theta \right ) } 
\right )
{\partial \delta \over \partial a} 
- {3 \over 2\, a^2 \left (1 \! + \! a^3 \, \theta \right ) } 
\left ( 
1 \! + \!  c_a { a^{3\! + \! \gamma_\nu} \, \theta \over 1 \! + \! a^3 \, \theta }
\right ) \delta = 0 
\label{eq:delta_a_eq_newt}
\eeq
for a reference scale $a_0=1$; the latter can always be re-introduced later by the 
trivial replacement $a \rightarrow a/a_0$.

Without a scale-dependent $G$ ($c_a=0$ in Eq.~(\ref{eq:grun_newt_a})), the growing
solution to the above equation is given by 
\beq
\delta_0 (a) \, \propto \,
a \cdot {}_2 F_1 \, \left ( {1 \over 3}, 1; {11 \over 6}; - a^3 \, \theta \right )  
\eeq
where  ${}_2 F_1$ is the  Gauss hypergeometric function.
To evaluate the correction to $\delta_0 (a) $ coming from the terms proportional to
$c_a$ one sets
\beq
\delta (a)  \propto \, a  \cdot {}_2 F_1 \, \left ( {1 \over 3}, 1; {11 \over 6}; -
a^3 \, \theta \right )  \, \left [ \, 1 + c_a \, {\cal F} (a) \, \right ] \; ,
\eeq
then inserts the resulting expression in Eq.~(\ref{eq:delta_a_eq_newt}), and 
finally expands the resulting expression 
to lowest order in $c_a$ to find the correction ${\cal F} (a)$.
The resulting differential equation can then be solved for ${\cal F} (a)$,
giving the density contrast $\delta (a)$ as a function of the 
two parameters ($\gamma_\nu$ and $\Omega$ or 
$\theta \equiv (1-\Omega)/\Omega$) appearing in Eq.~(\ref{eq:delta_a_eq_newt}).
In the following we will focus on the specific choice $\nu=\third$
obtained from the lattice theory of gravity \cite{ham00}, which leads
to the $G(a)$ exponent $ \gamma_\nu = {3 \over 2 \, \nu} = 9/2 $.
It is customary at this point to define the growth index 
$ f (a) \equiv { \partial \ln \delta (a) \over \partial \ln a } $
and the related growth index parameter $ \gamma $ via
$ \gamma \equiv  \left. { \ln f \over \ln \Omega  }  \right \vert_{a=a_0} $.
Then the solution to Eq.~(\ref{eq:delta_a_eq_newt}) gives an explicit 
expression for the growth index $\gamma$ parameter, as a function
of the matter fraction $\Omega$.

Based on observational constraints, one is mostly interested in the case $\Omega \approx 0.25$,
therefore in the following we will limit our discussion to this choice only.
In the absence of a running $G$ ($ G \rightarrow G_0 $, thus $c_a=0$) one has $ f (a=a_0) = 0.4625 $ 
and $ \gamma = 0.5562 $ for $ \Omega = 0.25 $ \cite{pee93}.
On the other hand when the running of $G$ is taken into account one finds from the solution to 
Eq.~(\ref{eq:delta_a_eq_newt}) for the growth index parameter $ \gamma $ at $\Omega =0.25 $
\beq 
\gamma =  0.5562 - 0.0142 \; c_a + O(c_a^2 ) \; .
\label{eq:gamma_newt_1}
\eeq
In the end it would seem therefore that at least in the Newtonian treatment the correction 
comes out rather small.
Note that both the Newtonian and the relativistic treatment, described in the main body, give
a negative sign for the correction arising from the running of $G$.

To estimate quantitatively the actual size of the correction in 
Eq.~(\ref{eq:gamma_newt_1}) one needs an estimate for the coefficient  
$c_0 \approx 33.3$ in Eq.~(\ref{eq:grun_box}),
as obtained from the lattice gravity calculations of invariant correlation functions at fixed geodesic distance \cite{cor94}.
In addition one uses the fact that $c_a \approx c_t \approx 2.7 \, c_0$
(see the previous discussion related to Eq.~(\ref{eq:grun_newt_a})).
From this one would then get the estimate $\gamma =  0.5562 - 1.28 $ on the largest scales,
which looks like a significant $O(1)$ correction to $\gamma$.

\vfill

\newpage

\end{document}